\documentclass[twocolumn]{revtex4}
\usepackage{natbib}
\usepackage{t1enc}
\usepackage{color}
\usepackage[dvips]{graphicx}
\usepackage{graphicx}
\usepackage{amssymb,amsfonts,amsmath}
\usepackage{ifpdf}
\pdfpagewidth\paperwidth 
\pdfpageheight\paperheight
\ifpdf
\fi

\begin{document}

\title{The zipper mechanism in phagocytosis: energetic requirements and variability in phagocytic cup shape}
\author{Sylvain Tollis$^{1, 2,}$, Anna. E. Dart$^{2, 3}$, George Tzircotis$^{2, 3}$, and Robert G.~Endres$^{1, 2, *}$}

\affiliation{
   {${}^1$Division of Molecular Biosciences}\\
   {${}^2$Centre for Integrative Systems Biology at Imperial College (CISBIC)}\\
   {${}^3$Division of Cell and Molecular Biology South Kensington Campus} \\
   {Imperial College London, SW72AZ London, United Kingdom} \\
   {$^*$ To whom correspondence should be addressed. E-mail: r.endres@imperial.ac.uk}}

\begin{abstract}
 
Phagocytosis is the fundamental cellular process by which eukaryotic cells bind and engulf particles by their cell membrane. Particle engulfment involves particle recognition by cell-surface receptors, signaling and remodeling of the actin cytoskeleton to guide the membrane around the particle
in a zipper-like fashion. Despite the signaling complexity, phagocytosis also depends strongly on biophysical parameters, such as particle shape, and the need for actin-driven force generation remains poorly understood.
Here, we propose a novel, three-dimensional and stochastic biophysical model of phagocytosis, and study the engulfment of particles of various sizes and shapes, including spiral and rod-shaped particles reminiscent of bacteria. Highly curved shapes are not taken up, in line with recent experimental results. Furthermore, we surprisingly find that even without actin-driven force generation, engulfment proceeds in a large regime of parameter values, albeit more slowly and with highly variable phagocytic cups. We experimentally confirm these predictions using fibroblasts, transfected with immunoreceptor Fc$\gamma$RIIa for engulfment of immunoglobulin G-opsonized particles. Specifically, we compare the wild-type receptor with a mutant receptor, unable to signal to the actin cytoskeleton. Based on the reconstruction of phagocytic cups from imaging data, we indeed show that cells are able to engulf small particles even without support from biological actin-driven processes.
This suggests that biochemical pathways render the evolutionary ancient process of phagocytic highly robust, allowing cells to engulf even very large particles. The particle-shape dependence of phagocytosis makes a systematic investigation of host-pathogen interactions and an efficient design of a vehicle for drug delivery possible.

\end{abstract}
\maketitle

\newpage

\section*{Background}
Phagocytosis is the ancient, evolutionarily conserved process by which eukaryotic cells bind, engulf, and destroy particles and cells larger than 0.5$\mu$m in diameter \cite{swanson2004,swanson2008,underhill2002}. The importance of phagocytosis is derived from its two main functions: (1) a feeding mechanism in single-cell organisms \cite{gerish2009}, and (2) the clearance of pathogens, apoptotic and senescent cells from our body by immune cells \cite{conner2003,greenberg2002}. As part of our immune defense, phagocytosis is mainly performed by professional phagocytes, including macrophages, neutrophils, and dendritic cells. Initiation of phagocytosis occurs with recognition of the target particle either directly or via an opsonising molecule. For instance the Fc portion of immunoglobulin G (IgG) is recognized by the cell-surface receptor Fc$\gamma$RIIa \cite{dupuy2008,groves2008}. Ligand-receptor binding triggers intracellular signaling \cite{groves2008,underhill2002,goldstein2001}, resulting in remodeling of the actin cytoskeleton \cite{castellano2001,may2001} and coherent growth of cell membrane around the particle to form the phagocytic cup \cite{swanson2004,swanson2008}. Eventually, the leading edge of the growing cup closes, and a membrane vesicle enclosing the particle (phagosome) moves inside the cell. Subsequently, the phagosome fuses with vesicles containing enzymes \cite{silva2007,yu2008}, acids \cite{kinchen2008}, and oxygen radicals \cite{hampton1998,segal2005} to destroy the particle.
\newline
\indent
\\
The biochemical pathways involved in phagocytosis are complex. Dozens of cell-surface receptors contribute to the recognition of a large variety of ligand molecules and subsequent particle engulfment \cite{groves2008,kraft2007,underhill2002}. The Fc$\gamma$ receptor (Fc$\gamma$R) \cite{daeron1997} and complement receptor 3 (CR3) of the integrin receptor family \cite{tohyama2008} are the most widely studied and understood receptors involved in phagocytosis. Fc$\gamma$R-mediated phagocytosis proceeds through membrane protrusions and leads to thin cups \cite{champion2006,niedergang2003}, whereas in CR3-mediated phagocytosis, particles appear to sink into the cell \cite{aderem1996,lecabec2002}. Spreading of the cell membrane over the particle involves actin-driven cell-shape changes similar to the processes involved in cell migration and adhesion \cite{dupuy2008,chan2009,colombelli2009,svitkina1999}. Specifically for Fc$\gamma$R, binding to an IgG-opsonized particle results in receptor clustering and recruitment of small GTPases of the Rho family, which, via proteins of the WASP family, activate the Arp2/3 complex \cite{swanson2004,underhill2002}. The latter promotes branching of actin filaments, leading to an increase in the number of uncapped ends and to an isotropic growth of the actin network \cite{groves2008,medalia2007}. Additionally, the phagocytic cup has been shown to be enriched in gelsolin \cite{yin1981,arora2005,serrander2000}, coronin \cite{gerish2009}, and other regulators of actin polymerization. All in all, this complex signaling pathway involves 100-1000 different types of molecules \cite{underhill2002,boulais2010}, rendering mathematical modeling at the molecular level impossible.
\newline
\indent
\\
Despite the huge biochemical complexity, the engulfment process shows a strong dependence on simple biophysical parameters. First, it relies on the availability of extra membrane at the phagocytic cup \cite{nelson1998,swanson1992}, provided by delivery of membrane vesicles \cite{niedergang2004} or unwrinkling of membrane folds \cite{hallett2007,herant2005}. Second, completion of phagocytic uptake depends on the shape of the particle and, interestingly, on the initial orientation of the particle on the cell surface \cite{champion2006,champion2009}. For instance, experiments demonstrate that elongated spheroid polystyrene particles coated with IgG are more efficiently engulfed when presented to the phagocyte with their tip first. Third, a recent study by one of the authors demonstrates that the biophysical requirements for phagocytosis lead to either complete phagocytosis or stalled cups due to the presence of a mechanical bottleneck \cite{tzircotis2009}. Interestingly, the same study shows that engulfment appears to even proceed in cells treated with (modest amounts of) cytochalasin D, an inhibitor of actin polymerization, indicating that biochemical pathways may not always be necessary for this initial stage of phagocytosis.
\newline
\indent
\\
The mechanism of phagocytosis is only partially understood, with key insights provided more than three decades ago. In the 1970's, Griffin and his collaborators \cite{griffin1975,griffin1976} demonstrated that incomplete coating of particles with ligand results in only partial uptake. This indicated that phagocytic uptake occurs via successive zipper-like ligand-receptor binding (Figure 1A), and not by an all-or-nothing mechanism triggered at the onset of phagocytosis. The zipper mechanism is the underlying assumption in a number of recent modeling works in phagocytosis \cite{herant2006,tzircotis2009} and endocytosis \cite{decuzzi2007,decuzzi2008,gao2005}, mainly addressing the influence of the cell-membrane tension and ligand-receptor bond density on engulfment. Despite the general acceptance of the zipper mechanism, many of its biophysical requirements are insufficiently understood. Questions, so far unanswered, include what the energetic requirements of the zipper mechanism are, specifically what role actin polymerization plays in its progression during phagocytosis, and also whether the zipper mechanism can explain the particle-shape dependence of phagocytosis. Previous models were unable to fully address the particle-shape dependence, as they assume rotational symmetry around the axis connecting cell and particle. Additionally, large particle-to-particle variation in cup growth \cite{tzircotis2009} and cell-to-cell variation in the related process of endocytosis \cite{snijder2009} point towards the importance of stochasticity during the uptake, not captured in previous deterministic approaches.
\newline
\indent
\\
Recent experiments provide new insights into the biophysical mechanism for driving the membrane around the particle, suggesting a ratchet-type mechanism. Once started, phagocytosis progresses unidirectionally and irreversibly \cite{herant2005}. This irreversible membrane progression is further supported by the loss of lipid and protein mobility at the phagocytic cup, observed using fluorescence recovery after photobleaching (FRAP) \cite{frap}. While several models proposed mechanisms of force generation by actin polymerization (see \cite{forcesbehind} and references therein), recent experiments based on fluorescent speckle microscopy demonstrate that actin does not directly push the membrane outwards. Instead, by filling gaps provided by membrane fluctuations (or other types of membrane movement), actin polymerization prevents the membrane from moving backwards like a ratchet \cite{herant2006}. The relevance of such Brownian ratchets in biology has previously been emphasized \cite{oster1,mogilner2003}. The question is if a ratchet mechanism, together with energetic restrictions in membrane bending and stretching, can naturally lead to phagocytic uptake and account for the shape-dependence of phagocytosis. 
\newline
\indent
\\
\begin{figure*}[!ht]
  \centering
   \includegraphics[]{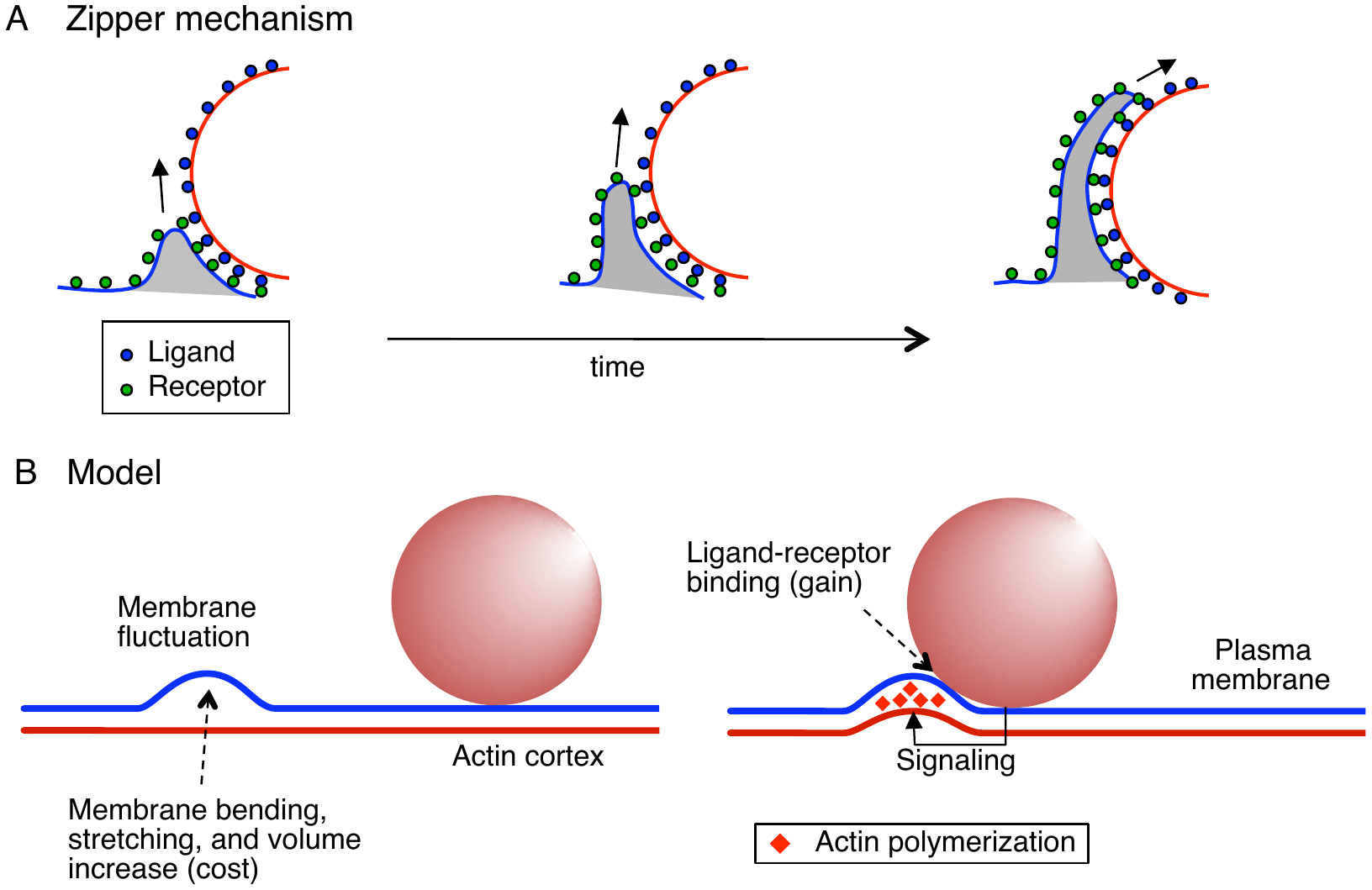}
   \label{fig01}
   \caption{ (A) Schematic of the zipper mechanism. Cell-membrane receptors (green dots) and ligands on the particle surface (blue dots) are sequentially engaged in bond formation, resulting in progression of engulfment with time (from {\it left} to {\it right}). (B) Schematic of our ratchet model. ({\it left}) A random membrane fluctuation (blue) far from the particle is unable to trigger ligand-receptor binding and signaling. Therefore it is not supported by remodeling of the actin cortex (straight red line), and the membrane may move backwards at a later time. ({\it right}) A membrane fluctuation near the particle leads to ligand-receptor binding, resulting in signaling and actin polymerization (red diamonds). Consequently, the actin cortex is deformed to support the membrane fluctuation, which makes the membrane move irreversibly for zipper progression. Energetic costs and gains of membrane fluctuations used in model are shown as well.}
\end{figure*}
In this work, we propose a ratchet-like biophysical model for the zipper mechanism. This model differs from previous works in that it is, to our knowledge, the first fully three-dimensional stochastic model of phagocytic engulfment.  Specifically, thermal membrane fluctuations, assumed to play a major role in our model, provide the energy source to locally deform the membrane and to build further ligand-receptor bonds for zippering the membrane around the particle. Actin polymerization makes ligand-receptor bonds effectively {\it irreversible}, {\it i.e.} reinforced and stabilized for a significant amount of time. To investigate the role of actin, we compare cup progression for the regular {\it active} zipper with a {\it passive} zipper model in which ligand-receptor binding remains specific and strong but {\it reversible} due to the absence of actin polymerization. 
\newline
\indent
\\
Interestingly, we find that the passive zipper also leads to engulfment of small particles, rendering phagocytosis highly robust. However, such passive engulfment is generally slower and produces much more variable phagocytic cups than the active zipper. Furthermore, our computer simulations lead to successful phagocytic engulfment in a broad range of parameters values, including different particle sizes. For non-spherical particles, completion of engulfment depends strongly on particle shape and orientation. Our model further predicts that cup shape invariably depends on membrane biophysical parameters, in particular surface tension and cell-volume constraint. 
\newline
\indent
\\
To test the predicted difference between the active and passive zippers, we experimentally implement the two different types of zippers using COS-7 fibroblasts which, after transfection with GFP-tagged Fc$\gamma$ receptor, phagocytoze IgG-coated polystyrene particles. Specifically, we performed phagocytic assays under three different {\it conditions}: (1) cells expressing wild-type Fc$\gamma$R for the active zipper (WT-Fc$\gamma$R), (2) cells expressing a signaling-dead mutant receptor (Y282F/Y298F-Fc$\gamma$R), which specifically binds IgG ligand but is unable to signal to the actin cytoskeleton \cite{mitchell1994,odin1991,tzircotis2009}, and (3) cells expressing WT-Fc$\gamma$R and treated with cytochalasin D (WT-Fc$\gamma$R$+$CytoD). The last two conditions represent two versions of the passive zipper due to the absence of actin polymerization in phagocytic cups. To compare with our model, we systematically analyze confocal microscopy images, and quantitatively estimate cup variability for the three different conditions using {\it small} (1.5 $\mu m$ radius) and {\it large} (3 $\mu$m radius) particles. Consistently with our simulations, phagocytic cups develop more slowly and are significantly more variable in the absence of actin polymerization. Our results provide new insights into the robustness of phagocytosis, as well as the role of bacterial cell shape in host-pathogen interactions.

\section*{Results and Discussion}
\subsection*{Ratchet model for the zipper mechanism}
\begin{figure*}
  \centering
    \includegraphics[]{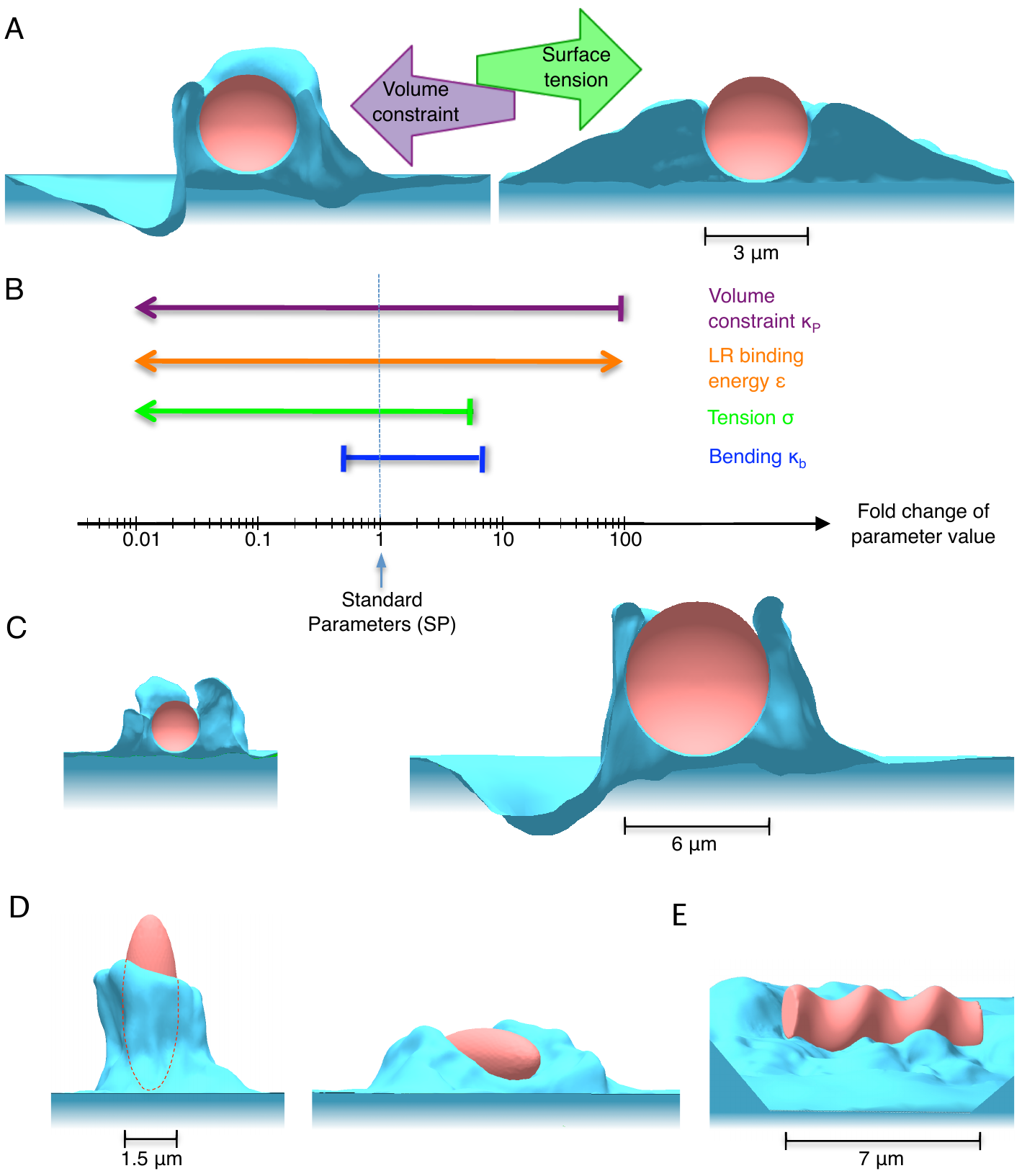}
\label{fig02} 
\caption{(A) Influence of surface tension and volume constraint. Simulated phagocytic cup for ({\it left}) a large volume constraint ($\kappa_{P}=10^{-2}$ $ \text{pN}\text{}\mu \text{m}^{-5}$) and a relatively low surface tension ($\sigma=6.2 \times 10^{-7}$ $\text{mNm}^{-1}$), and ({\it right}) a high surface tension ($\sigma=6.2 \times 10^{-5}$  $\text{mNm}^{-1}$) and a nearly unconstrained cell volume ($\kappa_{P}=3.8 \times 10^{-7}$ $ \text{pN}\text{}\mu \text{m}^{-5}$). Both sets of parameters produce approximately the same speed for engulfment. (B) Ranges of parameter values for successful engulfment. Simulation time was restricted to twice the engulfment time with Standard Parameters (SP) given in {\it Methods} and represented by vertial dashed line. Shown are fold changes of bending stiffness (blue), surface tension (green), ligand-receptor energy density (orange) and volume constraint (purple) along x-axis. Pointed arrows indicate that parameter regimes may extend beyond tested limits, blunt arrows indicate sharp limits beyond which full engulfment is not reached in simulation time. (C) Cross section of particle engulfment with particle radii $R=1.2 \mu \text{m}$ ({\it left}) and $R=3.8 \mu \text{m}$ ({\it right}). (D) Engulfment of a spheroidal particle with tip ({\it left}) and long side ({\it right}) first. Principal axis of the spheroidal particle are $R_1=R_2=1.5 \mu \text{m}$, and $R_3=4.2 \mu \text{m}$. Dashed curve indicates particle outline. (E) Stalled engulfment of spiral-shaped particle, characterised by a volume similar to a spherical particle of radius $2.2\mu\text{m}$.}
\end{figure*}
Our model is based on the following experimental observations. Engulfment of quasi-spherical particles by neutrophils progresses continuously without significant pause or reversal, indicating that ligand-receptor binding is essentially irreversible \cite{herant2005}. This irreversibility is further supported by FRAP and single-molecule experiments, which show that lipids and proteins in phagocytic cups, as well as ligand-bound Fc-receptors are immobilized in an actin-dependent fashion \cite{frap,andrews2008}. Additional support for the notion of irreversible uptake was recently determined in a related context \cite{speckle}. Fluorescent speckle microscopy of actin flow and image analysis during cell migration show that the membrane at the leading edge protrudes first, followed by actin polymerization to fill the gap between the membrane and the actin cortex. Such actin polymerization is mainly restricted to the leading edge due to signaling by receptors and/or localization of small GTPases of the Rho family \cite{groves2008}. The role of actin polymerization in phagocytosis is hence to stabilize ligand-receptor bonds and to rectify membrane movements in a ratchet-like fashion, leading to unidirectional movement of the leading edge of the engulfing cell.
\newline
\indent 
\\
Figure 1A introduces the general concept of the zipper mechanism in phagocytosis, and Figure 1B summarizes our ratchet model for this mechanism. The cell membrane and the actin cortex are described by a Helfrich-type energy function \cite{helfrich1973}, including contributions from ligand-receptor binding, membrane bending and stretching, as well as a cell-volume constraint. Chosen membrane parameters effectively describe the cell plasma membrane with its underlying actin cortex. The model was implemented using finite-temperature Monte Carlo simulations of the discretized cell membrane (see {\it Methods} for details). Briefly, the algorithm proposes random, thermally generated membrane fluctuations (trial moves), which are either accepted or rejected depending on the change in total energy during the move. When a random membrane fluctuation brings the cell membrane in contact with the particle, this fluctuation is likely accepted due to energetically favorable ligand-receptor binding. Once accepted, this fluctuation is made irreversible as a result of signaling and actin polymerization. In contrast, a membrane fluctuation far away from the particle is less likely to be accepted. Even if accepted, the fluctuation is not made irreversible and hence may retract at a later time (see Supplementary Fig. S1). Hence, in our model the actin network only supports membrane fluctuations which lead to progression of the engulfing zipper as a result of signaling.
\newline
\indent

\subsection*{Dependence of phagocytic cup shape on membrane biophysical parameters}
\begin{figure*}[!ht]
 \centering
    \includegraphics[]{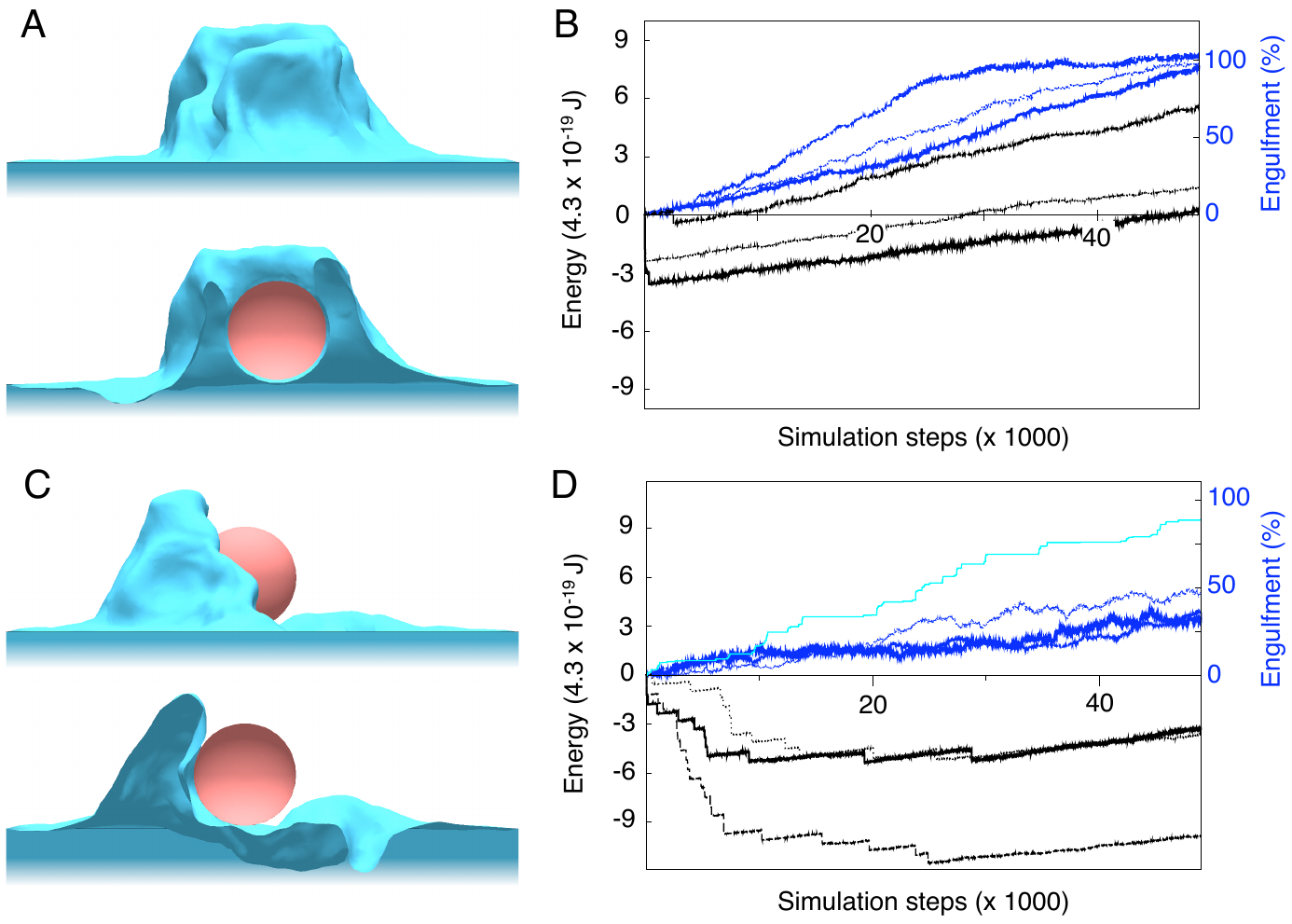}
\label{fig03}
\caption{(A) Side view (top) and cross section (bottom) of a phagocytic cup obtained for the active zipper. (B) Corresponding time courses of the membrane energy (thick solid, medium dashed and thin dotted  black lines) and percentage of engulfment, defined by the average membrane height around the particle (blue solid, blue dashed, blue dotted lines) for three repeats of the stochastic simulation. (C) Side view (top) and cross section (bottom) of the phagocytic cup obtained for the passive zipper (same overall simulation time). (D) Corresponding time courses of the membrane energy (thick solid, medium dashed and thin dotted  black lines) and engulfment (blue solid, blue dashed, blue dotted lines) for three repeats of the simulation. Dashed light blue line indicates corresponding maximal membrane height for one of the simulations (in percentage of the particle diameter). Small particles with $1.5\mu$m radius were used.}
\end{figure*}
Using our model for the zipper mechanism, we have successfully simulated phagocytic engulfment in a broad range of parameter values (see Figure 2). Figure 2A shows two different characteristic cup shapes we obtained. Low surface tension ({\it i.e.} low energy cost for stretching the membrane and underlying actin cortex), and tight cell-volume constraint ({\it i.e.} high energy cost for increasing the cell volume), lead to a thin phagocytic cup since a thin cup requires extra membrane but little extra volume. In contrast, weak volume constraint and/or high surface tension produce a broad cup. Based on the parameters explored, we chose intermediate values for both surface tension and cell-volume constraint as our Standard Parameters (SP) for the remainder of our simulations in order to produce realistic cup shapes (see {\it Methods} for details). Figure 2B shows that most parameter values can be changed independently by at least one order of magnitude, without negatively affecting engulfment completion. Note that changing simultaneously several parameters may affect engulfment more drastically. Our simulations also show that cup shape depends on the kinetics of engulfment, determined by membrane fluctuations and therefore temperature (see Supplementrary Fig. S2). Additionally, preventing thermal fluctuations (by setting the temperature to zero Kelvin) during a simulation stops cup progression. This indicates that in our model membrane fluctuations are indeed required to bring receptors in close contact with ligand molecules on the particle, emphasizing their important role in the ratchet mechanism.

\subsection*{Active versus passive uptake and the role of actin}
Although phagocytosis generally involves active processes such as actin polymerization in the cup (active zipper) \cite{may2001}, recent reports indicate that phagocytosis may still work in an actin-independent manner. Indeed, phagocytic uptake was observed despite treating phagocytes with (modest amounts of) cytochalasin D \cite{tzircotis2009}. Hence, ligand-receptor binding may be sufficient in guiding the cell membrane around the particle under certain conditions (passive zipper). To investigate the energetic requirements of the zipper mechanism, we implemented simulations of the passive zipper. In these simulations, ligand-receptor bonds are not stabilized by actin polymerization and can unbind at later times, {\it i.e.} remain reversible. Hence, engulfment may still progress if the energetic cost of stretching and deforming the membrane is offset by the ligand-receptor binding energy in the presence of thermal membrane fluctuations. 
\newline
\indent
\\
Figure 3 ({\it left}) shows that engulfment of small ($1.5\mu$m radius) particles by the passive zipper leads to more variable phagocytic cup shapes than engulfment by the active zipper. For the active zipper, random membrane fluctuations are rectified by irreversible ligand-receptor binding due to actin polymerization. This leads to uniform progression of the cell membrane all around the particle at approximately the same speed (Figure 3A). In contrast, engulfment by the passive zipper occurs through binding of large membrane ruffles which eventually enclose the particle (Figure 3C). The variability of the phagocytic cup may be a measure of the respective contributions of active and passive processes in engulfment progression. 
\newline
\indent
\\
Figure 3 ({\it right}) compares time courses of the membrane energy and progression of uptake for the active and passive zippers. We found that in both cases, membrane energy decreases rapidly at the very beginning of the uptake process due to energetically favorable ligand-receptor binding without large, energetically unfavorable deformations of the cell membrane. After this short initial period, the total energy increases with simulation time. This increase is much more pronounced for the active zipper, which stabilizes energetically unfavorable random membrane fluctuations by actin polymerization. In contrast, for the passive zipper such high-energy membrane deformation may not last over time. The slower increase in energy for the passive zipper correlates with a slower engulfment. For the simulations shown, engulfment for the active zipper is approximately twice to three times as fast as for the passive zipper, although the latter eventually engulfs the particle. However, the difference between active and passive engulfment depends on biophysical parameters, and may be reduced for particular choices of the parameters values, {\it e.g.} lower surface tension and/or stronger ligand-receptor binding.
\newline

\subsection*{Particle size matters for passive, not for active zipper}
\begin{figure*}
  \centering
    \includegraphics[]{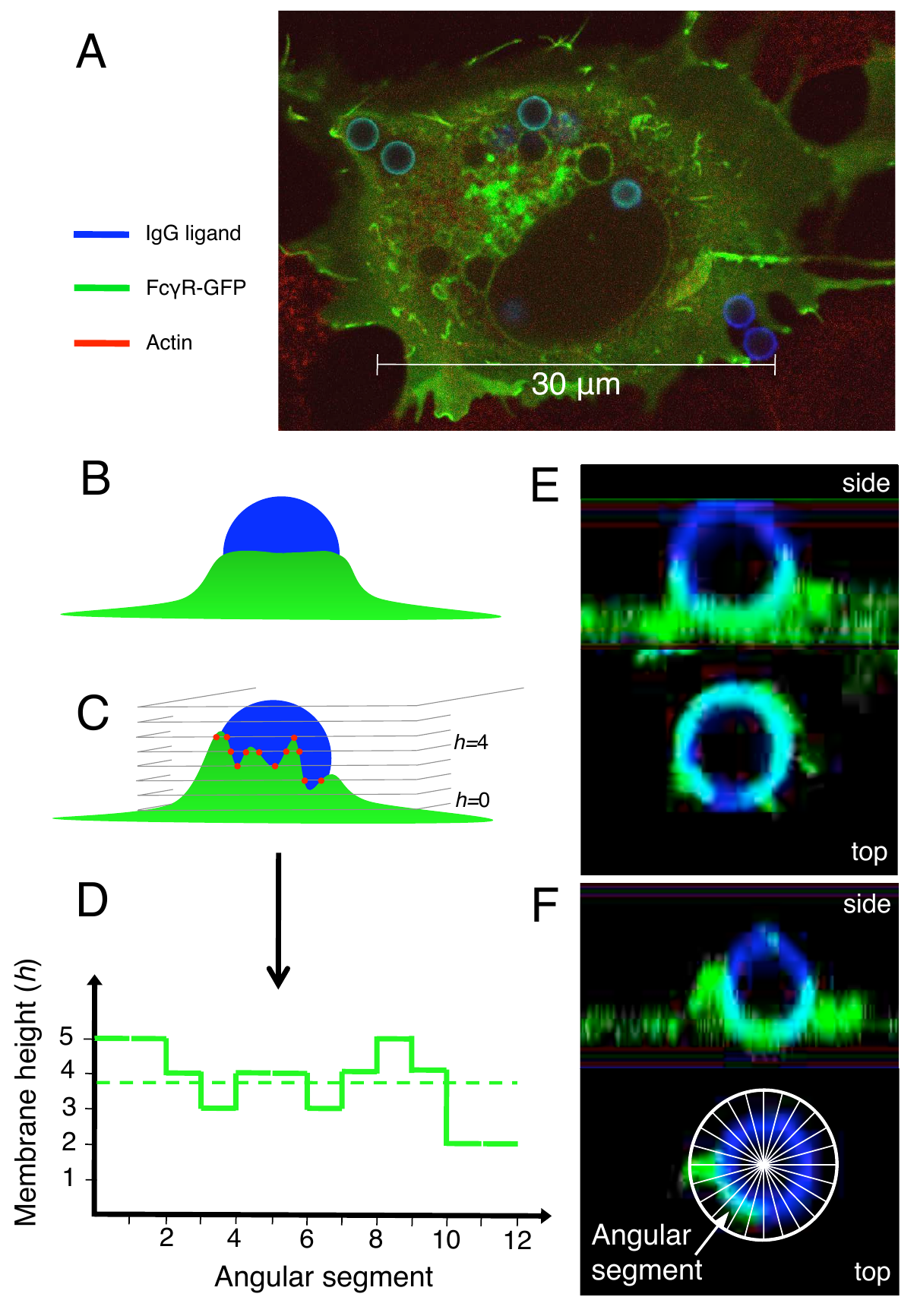}
\label{fig04}
\caption{(A) Typical fluorescence image of COS-7 cell with Fc$\gamma$R-GFP shown in green and IgG antibodies on $1.5\mu$m-radius particles shown in blue. (B, C) Illustrative schematics of 3-dimensional confocal fluorescence microscopy images. (B) Schematic (side view) of a regular cup. (C) Schematic (side view) of a variable cup. From the intersections of the confocal imaging planes with the phagocytic cup, the maximal height of the cell membrane is determined for each angular segment around the particle (red dots). (D) Membrane height as a function of angular segment (see inset in panel F), corresponding to the cup drawn in panel B (dashed line) and C (solid line). Only the angular segments from index 1 to 12 out of 24 are shown. The variability of the cup is defined as the standard deviation divided by the average of this function. The higher this measure, the more variable the cup. (E, F) Side and top views of half-engulfed particles, with cups reconstructed from confocal microscopy data. (E) Top view, taken in the equator plane of the particle, shows a regular distribution of cell membrane (receptors) around the particle (regular cup). (F) Top view shows an irregular distribution of cell membrane (variable cup). {\it Inset} Schematic defines the angular segments (in top view).}
\end{figure*}
Experiments show that phagocytosis is relatively insensitive to particle size \cite{herant2006}. Using our model for the active zipper, we simulated engulfment of spherical particles with different radii ranging from $1.2$ to $3.8\mu$m. Figure 2C shows that engulfment progresses normally for small and large particles. Hence the active zipper mechanism is sufficiently robust to allow engulfment of differently sized particles using the same set of biophysical parameters, although engulfment of large particles requires more time. Noticeably, large particles (in general, with radius larger than $2.5\mu$m) were taken up via more regular phagocytic cups than small particles (with radius $1.5\mu$m or smaller), indicating that active processes may be required for engulfment of large particles. To confirm this observation we have simulated engulfment of large $3\mu$m-radius particles by both the active and the passive zipper, shown in Supplementary Fig. S5. While the active zipper resulted in complete uptake of the particle, the passive zipper only engulfed a few percent of the particle's surface area. Thus the difference in phagocytic efficiency between the two zipper types was exacerbated for large particles, reflecting the importance of actin polymerization for engulfment of large particles.
\newline

\subsection*{Experimental test of model predictions}

To test our model predictions and to specifically compare active with passive engulfment for small and large particles, we transfected COS-7 cells with either wild-type Fc$\gamma$R (WT-Fc$\gamma$R) or a signaling-dead mutant receptor (Y282F/\ Y298F-Fc$\gamma$R). Cells expressing the wild-type receptor are expected to perform active engulfment, whereas cells expressing the signaling-dead mutant receptor are expected to perform passive engulfment. As a control, passive engulfment is additionally implemented by treating cells expressing WT-Fc$\gamma$R with $0.2\mu$M of cytochalasin D (WT-Fc$\gamma$R$+$CytoD), which prevents actin polymerization (see {\it Methods}). Synchronized phagocytosis assays using small (1.5 $\mu$m radius) and large (3 $\mu$m radius) IgG-opsonized polystyrene particles were carried out, and, after fixation, receptor localization in phagocytic cups was visualized by fluorescence confocal microscopy. Cells were imaged at different time points during phagocytosis, and at each time point, three to eight imaged cells were each engulfing simultaneously between four and twenty particles (Figure 4A). Consequently, for each condition we analyzed at least seventy particles.
\newline
\indent
\\
\begin{figure*}[!ht]
  \centering
    \includegraphics[]{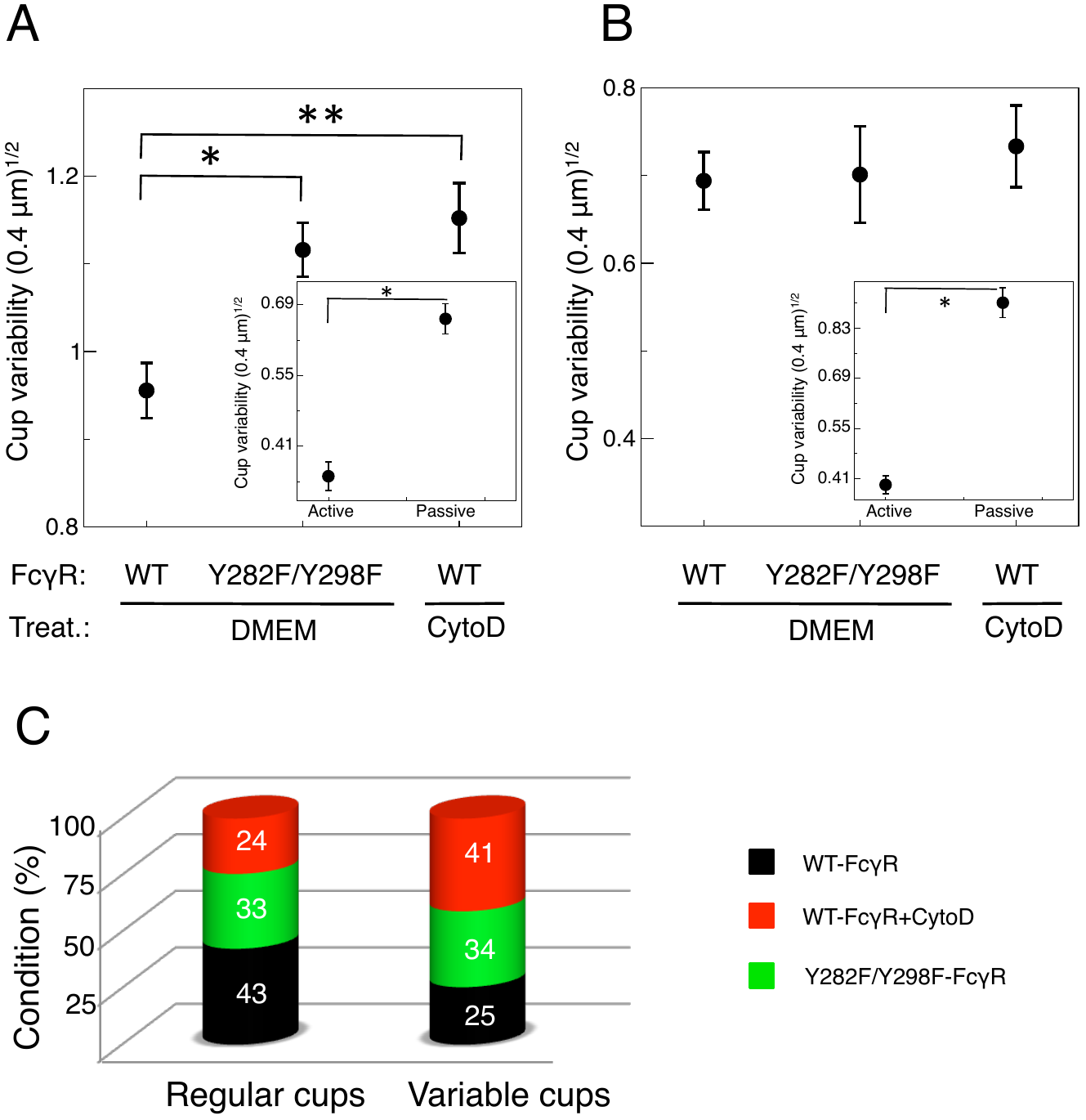}
\label{fig05}
\caption{(A) Experimental cup variability, defined in Figure 4, for 20-40\% engulfed particles, for COS-7 cells transfected with wild-type Fc$\gamma$R (WT-Fc$\gamma$R) or signaling-dead mutant (Y282F/Y298F-Fc$\gamma$R) in DMEM medium, as well as WT-Fc$\gamma$R cells treated with $0.2\mu$M cytochalasin D (WT-Fc$\gamma$R$+$CytoD). The data points represent the variability averaged over all the phagocytic cups measured for given condition (receptor type and treatment) and range of engulfment. Error bars represent the statistical standard error of the cup variability. Student's t-test: p-values are (*) 0.0002 and (**) 0.0003. ({\it Inset}) Theoretical cup variability calculated from corresponding model simulations. Data points and error bars represent respectively the statistical average and standard error of the cup variability over several simulations. Student's t-test (*): p-value is given by 0.0002. (B) Experimental cup variability for 40-60\% engulfed particles. ({\it Inset}) Theoretical cup variability calculated from corresponding model simulations. Student's t-test (*): p-value is given by 0.0001. (C) Repartitioning of regular and variable cups for different experimental conditions, {\it i.e.} wild-type receptor (WT-Fc$\gamma$R), signaling-dead mutant (Y282F/Y298F-Fc$\gamma$R) and cytochalasin-D treated cells (WT-Fc$\gamma$R$+$CytoD), for engulfment between 20 and 40\%. The numbers indicate the contribution (in \%) of the three conditions in the overall sample of regular ({\it left}) and variable ({\it right}) cups.}
\end{figure*}
To test whether passive engulfment leads to more variable cups than active engulfment, we developed an image-analysis method illustrated in Figure 4B-F. The cup shape varibility was quantified by the standard deviation of the distribution of cell-membrane (Fc$\gamma$R-GFP fluorescence) height around the particle, divided by the square root of the average membrane height. The unit of membrane height is given by the distance ($0.4\mu$m) between consecutive confocal image planes (see {\it Methods}). Figure 5A shows that for small particles engulfed between 20 and 40\% of their surfaces, cup variability increases from cells transfected with wild-type receptor to cells transfected with signaling-dead mutant receptor to WT-Fc$\gamma$+CytoD cells. The lowest variability, found for cells expressing wild-type receptor, is statistically significant against both passive zipper types (Student's t-test, p-value $<0.001$). This result is consistent with model predictions: Figure 5A, {\it inset} shows the cup variability from simulations, revealing that the active zipper leads to significantly less variable cups. In contrast, for the ranges of engulfment between 40 and 60\% (Figure 5B) and between 60 and 100\% (see Supplementary Fig. S7) we observed no noticeable difference in cup variability between the three experimental conditions, while our model consistently predicts more variable cups for the passive zipper (Figure 5B, {\it inset}). This discrepancy may indicate that active processes such as contraction by myosin motor proteins become important at later stages of engulfment, limiting our model's full validity to the early events in phagocytosis (see {\it Conclusion} section). 
\newline
\indent
\\
To confirm that our results are independent of the specifics of the analysis method used, we also analyzed phagocytic cups with an alternative, albeit less accurate, method (see Supplementary Fig. S9). This method consists of determining the distribution of Fc$\gamma$R-GFP fluorescence intensity around the particle at its equator plane, restricting the analysis to approximately half taken up particles ($30-70\%$). The standard deviation of this distribution provides an alternative measure of the cup variability. We arrived at the same conclusion, confirming our result of the difference in cup variability (see Supplementary Fig. 10). Finally, note that temperature-induced synchronization is imperfect and may lead to variability in cup growth \cite{tzircotis2009}. However, our measures of the variability in cup shape are independent of such an effect since we include all time points together in the analysis (except for plots showing the time dependence of engulfment).
\newline
\indent
\\
Our experiments further show that cup shape ranges from regular to variable for all three experimental conditions, but that the frequency of different cup shapes depends on the condition. Figure 5C plots the repartition of phagocytic cups for different conditions into both regular and variable cups. Note that a cup was identified as regular if its variability was below the variability averaged over all experimental conditions. In contrast, a cup was identified as variable if its variability was above the overall average. This plot shows that in our experiments a regular cup is most likely produced by a cell expressing wild-type receptor, whereas a variable cup is most likely to be produced by a cytochalasin-D treated cell. Hence, the cup shape has universal features independent of biochemical details. Examples of a regular and a variable cup are provided in Figures 4E and F, respectively. Both cups were taken from a cytochalasin-D treated cell, confirming that a regular cup can occur under any of our experimental conditions. 
\newline
\indent
\\
Our model also predicts that uptake by the active zipper is significantly faster than with the passive zipper (see Figures 3B and D). We experimentally tested this prediction by determining the percentage of engulfed surface area for each particle for different time points after initiation of phagocytosis, and comparing this result with our simulations, in which simulation time was matched to actual time. Figure 6A shows that cells transfected with the wild-type receptor (active zipper) engulf significantly faster (three to four times) than cells under the other two conditions (passive zippers). This result is in quantitative accordance with our model predictions (Figure 6B). Furthermore, we determined the time dependence of phagocytic uptake for large particles. The active zipper, although slower for large than for small particles, still engulfs regularly, both in experiments (Figure 6C) and simulations (Figure 6D). Note that predicted and measured time courses are in very good agreement without rescaling the time axis of the large-particle simulation. Furthermore, Figures 6C and 6D demonstrate the inability of the passive zipper to take up large particles, in both experiments and simulations. After more than 10 minutes, the average engulfed surface area remains below 20\%.
\newline
\indent
\\
\begin{figure*}
 \centering
   \includegraphics[]{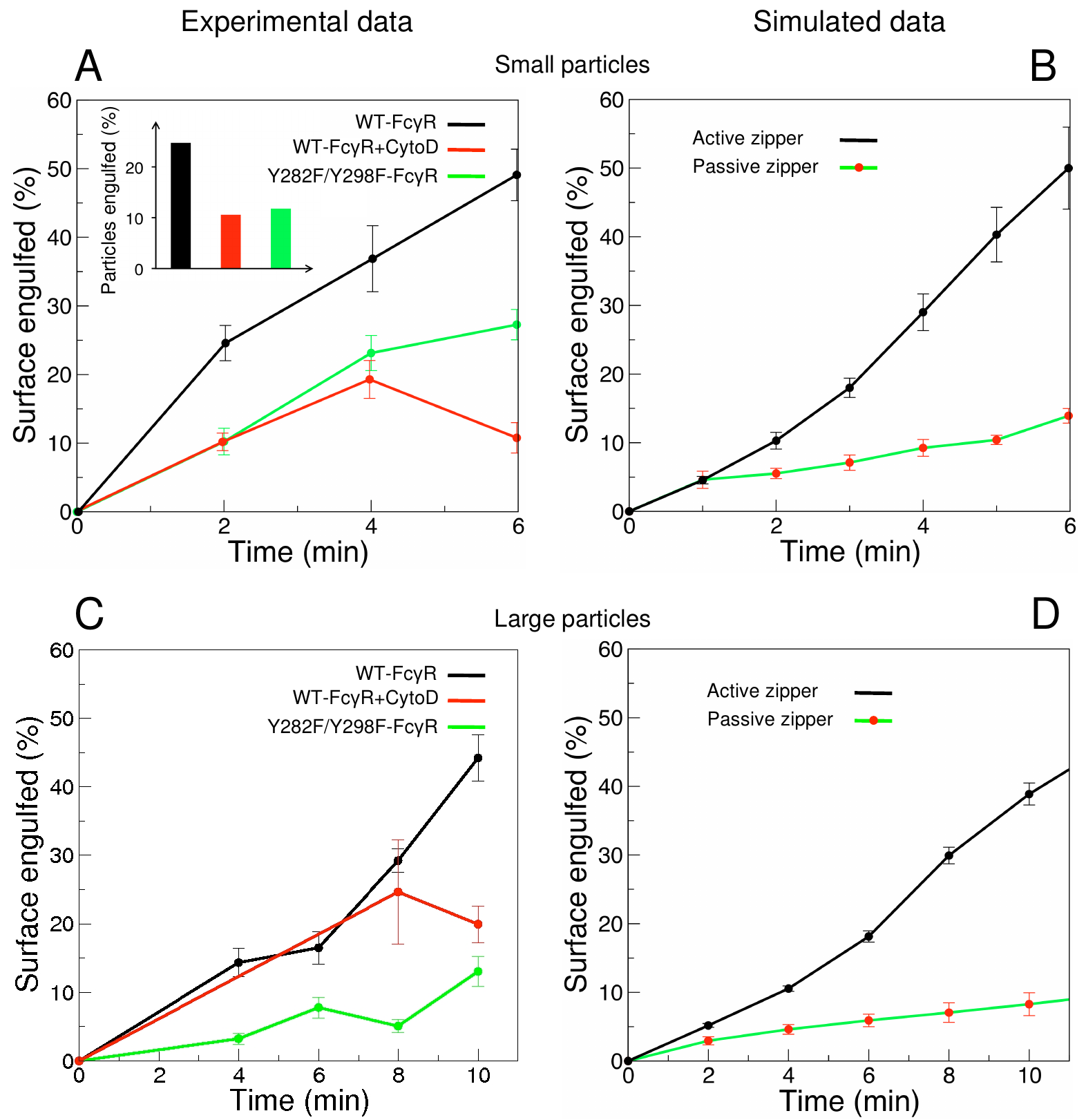}
\label{fig06}
\caption{Experimental data for small particles with $1.5\mu$m radius (A) and large particles with $3\mu$m radius (C). The data points represent averages of phagocytic cups obtained at given time points for the three different experimental conditions. Error bars represent the corresponding standard errors. (A, {\it Inset}) Proportion of almost completely taken up particles (engulfed surface area larger than 70\%) after at least 6 minutes for the three conditions. (B, D) Progression of phagocytic engulfment with time in the simulations for small (B) and large (D) particles. Simulation time was converted to actual time so that 50\% average engulfment of small particles by the active zipper corresponds to 6 min. No further adjustment of time conversion was done for large-particle simulations. The data points and error bars represent respectively the statistical average and standard error of phagocytic cups size from several simulations.}
\end{figure*}
Note that for the time points measured, the average uptake does not exceed 50\%, even for WT Fc$\gamma$R. This is caused by the fact that some particles are not engulfed irrespective of the condition \cite{tzircotis2009} (see Supplementary Fig. S6), reducing the average percentage of engulfment. The proportion of almost completely engulfed particles (with engulfed surface area larger than 70\%) beyond 6 minutes is represented in the inset of Figure 6A, showing that completion of engulfment is possible even for cells without actin polymerization. Note that long phagocytic assays (12, 14, and 45 minutes) were performed for cytochalasin-D treated cells only, explaining why the difference in complete engulfment with cells expressing the wild-type receptor is smaller than one may expect.
\newline
\indent
\\
From these results, we conclude that functional actin polymerization is required for fast and regular engulfment in phagocytosis. Nevertheless, in line with our model predictions, cells showing deficient actin polymerization at cups are still able to uptake small particles, although more slowly and with more variable cups.

\subsection*{Active zipper reproduces particle-shape dependence of phagocytosis}
Previously published experiments show that phagocytosis depends strongly on particle shape. In particular, elongated particles (similar to rod-shaped bacteria) are only taken up when presented to the phagocyte with their tip first \cite{champion2006,champion2009}. To test the particle-shape dependence of phagocytosis we conducted simulations of the active zipper using particles of different shapes while varying the initial orientation of the particle on the cell surface. Figure 2D shows the uptake of a prolate spheroid for two different orientations on the cell membrane after the same elapsed simulation time. In accordance with experimental observations \cite{champion2006,champion2009}, uptake is more advanced for the spheroid particle engulfed with its tip first (about 80\% engulfed surface area) than for the particle attached along its major axis (about 50\% engulfed in the same amount of time). This suggests a strong inhibitory effect of high local curvature on uptake. In our model this is readily attributed to the energetic cost of bending the membrane around the two highly curved ends of an elongated particle placed horizontally on the cell membrane. In line with this explanation and experiments \cite{champion2009}, the spiral-shaped particle in Figure 2E is not engulfed after a time duration sufficient for the engulfment of large spherical particle of twice its volume. Hence, our simulations demonstrate that particle shape and orientation are indeed important biophysical parameters for phagocytosis.

\section*{Conclusion}

In this work, we studied the biophysical requirements of the zipper mechanism, in particular the role of receptor-induced actin polymerization, and the effect of particle shape on uptake. In our model, the underlying biophysical mechanism of the zipper is an actin-driven thermal ratchet, which renders random membrane fluctuations irreversible close to the particle (Figure 1). This mechanism is supported by several recent experiments \cite{herant2005,frap,speckle}. Previously, such Brownian ratchet models were successfully applied to explain force generation by actin polymerization and motility of the pathogen {\it Listeria} in hosts cells \cite{oster1}. Our fully stochastic simulations can address for the first time the variability in particle uptake, recently noticed in phagocytic cup growth \cite{tzircotis2009} and completion of endocytosis \cite{snijder2009}. Implementation of our model in simulations led indeed to phagocytic engulfment for a broad range of values of membrane parameters (Figure 2), indicating exquisit robustness of the phagocytic process. However, phagocytic cup shape depends on parameter values, specifically on the ratio between surface tension and cell-volume constraint, as well as on the kinetics of engulfment. Cells with low surface tension and/or tight volume constraint develop thin cups (Figure 2A, {\it left}), characteristic of Fc$\gamma$R-mediated phagocytosis  \cite{aderem1996,gerish2009}. In contrast, cells with high surface tension produce broad cups (Figure 2A, {\it right}). The latter cup shape is more reminiscent of CR3-mediated phagocytosis, although for this type of phagocytosis particles are believed to sink into cells without protrusive cups \cite{aderem1996,lecabec2002}.
\newline
\indent
\newline
\indent
Using our model we were able to address the question whether the zipper mechanism requires an active driving force, such as provided by actin polymerization. For this purpose, we compared the regular active zipper with a passive version of the zipper. In the passive zipper, ligand-receptor bonds are as strong as for the active zipper (based on experimental observation \cite{tzircotis2009}) but are reversible, {\it i.e.} are not supported by actin polymerization. We demonstrated that the passive zipper also leads to engulfment of small particles (of radius $1.5\mu$m), although cup progression is slower and more variable (see Figure 3). Our active zipper can also reproduce the independence of uptake on particle size, in line with experimental observations \cite{herant2006}. In contrast, large particles (of radius $3\mu$m) are poorly phagocytozed by the passive zipper. We subsequently confirmed these predictions with experiments by transfecting COS-7 fibroblasts with wild-type Fc$\gamma$R and signaling-dead mutant Y282F/Y298F Fc$\gamma$R (see Figures 5 and 6). While the wild-type receptor represents the active zipper, the passive zipper is implemented through the use of signaling-dead mutant receptor or treatment with cytochalasin D. Both prevent actin polymerization in the cups. Our study may indicate that ancient forms of phagocytosis were driven by physical (passive) principles, and only later in evolution biochemical regulatory pathways were added for further support and robustness. Passive phagocytosis may also become important when energy sources are scarce.
\newline
\indent
\newline
\indent
Despite the robustness of phagocytosis to particle size, there appears to be a mechanical bottleneck around half-engulfment, recently observed by imaging \cite{tzircotis2009} and also predicted by our model. For slight variations in some of the parameters, our simulations of the active zipper produce either complete or significantly incomplete uptake (see Supplementary Fig. S6). Indeed, when the cup grows, deforming the membrane costs more and more energy per surface area engulfed. Beyond half-engulfment, the surface tension energy is twice as high as at the beginning of engulfment due to the membrane folding back onto itself. If this energy cannot be provided by the zipper, then cup progression stalls and the particle remains incompletely engulfed. Alternatively, the experimental data on incomplete particle uptake may be the result of particle attachment on cell-membrane areas, unable to phagocytose due to other reasons, such as unfavorable local cell-surface curvature, proximity to cell edge or nucleus, or missing proteins, lipids, and smaller molecules belonging to key signaling pathways. Further studies will be required, ideally using live-cell imaging to avoid the need for conserving the cell's internal structures by fixation \cite{dieckmann2010}.
\newline
\indent
\\
Our model for the zipper mechanism can also explain the strong particle-shape dependence observed in phagocytosis. Experiments show that an elongated spheroid is rapidly engulfed if the particle attaches to the cell membrane with its tip, but not if the particle attaches along its major axis \cite{champion2006,champion2009}. Furthermore, spiral-shaped particles are not phagocytozed \cite{champion2009}. The strong particle-shape dependence of phagocytosis is likely of biological relevance. On the one hand, it may increase the rate of infection of host cells by pathogenic bacteria. Indeed, recent experiments show that {\it Mycobacteria tuberculosis} and {\it marinum} are efficiently taken up and later released for spreading of the infection with the bacteria's tip first \cite{hagedorn2009}. On the other hand, the highly curved shapes of some bacteria, {\it e.g.} the spiral-shaped {\it Helicobacter} and {\it Campylobacter} species, may prevent their uptake by macrophages \cite{champion2009}, although injection of effector proteins can also be used by pathogens to hijack the immune or host cell's phagocytic response \cite{rosenberger2003,cornelis2002}. Furthermore, the particle-shape dependence of phagocytosis may be exploited to improve drug delivery by enclosing active drugs in particles, whose shape prevents uptake and destruction by macrophages \cite{champion2009,muttil2009}.
\newline
\indent
\\
While our biophysical model for the zipper mechanism is readily accessible for analysis and interpretation, the small number of model parameters makes a direct comparison with measured parameter values difficult. First, our membrane parameters such as surface tension and bending stiffness are an order of magnitude smaller than reported bulk membrane parameters (see {\it Methods}). This reduction is not surprising as cells regulate these parameters locally for efficient uptake \cite{charras2005,huang2005,roduit2008}. Such regulation may include lowering of the surface tension by local membrane delivery through vesicles \cite{niedergang2004,raucher1999,raucher1999-2} and unfolding of membrane wrinkles \cite{hallett2007}, as well as changes in the lipid and protein composition in the phagocytic cup \cite{swanson2004}. Second, our description of ligand-receptor interaction assumes that ligand and receptor distributions are continuous and homogeneous, while experiments indicate the formation of receptor micro-clusters \cite{sobota2005}, possibly as part of lipid rafts \cite{frap}. Third, our mechanism for actin polymerization based on membrane fluctuations neglects the role of the motor proteins (myosin I and II), whose role in membrane deformations has been established \cite{kress2007,swanson1999}. Consequently, our simulations describe well the dynamics of engulfment during the first two thirds of the uptake. At later stages of uptake, experiments show that phagocytic cups close rapidly with a thin membrane protrusion \cite{gerish2006,gerish2009}, while our simulations show slow cup closure. However, taking into account myosin-driven contraction is beyond the scope of the current work.
\newline
\indent
\\
Our model may also be applicable to other biological systems in which a zipper-like mechanism is involved. One such example is sporulation of {\it Bacillus subtilis} during starvation. After asymmetric cell division, the larger mother cell engulfs the smaller forespore for spore maturation. Interestingly, the mother cell even engulfs the forespore when the cell wall is artificially removed. This process occurs in a fast, zipper-like fashion without known sources of energy \cite{pogliano2006}. Importantly, forespore engulfment is subject to high variation. About 60\% of the cells successfully complete forespore engulfment, while 40\% do not at all, similar to the observation of the mechanical bottleneck in phagocytosis. Other examples of engulfment may not be driven by a zipper mechanism. For instance, the penetration of red blood cells by the malaria {\it Plasmodium} merozoite is driven by an elaborate actin machinery of the particle (the parasite), which devotes all its resources to wrap the cell membrane around itself. In stark contrast to phagocytosis, the engulfing host cell is completely passive \cite{malaria}.

\section*{Methods}
\subsection*{Theoretical techniques}
\noindent {\bf Cell membrane and particle models.} 
The cell membrane is described by a two dimensional elastic sheet \cite{brown2008,helfrich1973}, which includes both its lipid bilayer and associated actin cortex. The particle is assumed to be rigid and immobile. Moreover, we account for ligand-receptor binding and include a constraint on the cell volume. Therefore, the total free energy is given by
\begin{equation}
E=E_{\text{m}}+E_{\text{vol}}+E_{\text{LR}} \label{eq:free_energy_total} \text{,}
\end{equation}
where
\begin{equation}
E_{\text{m}}=\int_{m} d^{2}\mathbf{r}\left[\frac{\kappa_{b}}{2} C^2(\mathbf{r})+\sigma\right] \label{eq:free_energy_membrane}
\end{equation}
with $C^2(\mathbf{r})=C{_1}^{2}(\mathbf{r})+C{_2}^{2}(\mathbf{r})$ the square curvature obtained from the minimal ($C{_1}$) and maximal ($C{_2}$) curvatures at point $\mathbf{r}$. Note that the term corresponding to the product $C{_1}(\mathbf{r})C{_2}(\mathbf{r})$ is independent of the actual shape of the membrane as long as the overall topology is conserved, and therefore is ignored (Gauss-Bonnet theorem). Bending stiffness $\kappa_{b}$ reflects the energy cost of bending, and surface tension $\sigma$ reflects the energy cost of stretching the membrane with underlying actin cortex.
Furthermore, expanding or shrinking the cytosol locally costs the energy
\begin{equation}
E_{\text{vol}}=E_{\text{cell}}(V)-E_{\text{cell}}(V_0)=\kappa_{P}(V-V_0)^2 \label{eq:free_energy_volume} \text{,}
\end{equation}
where the quadratic dependence on actual volume $V$ comes from the lowest order Taylor expansion of the cell energy around local steady state volume $V_0$, and $\kappa_{P}=\frac{1}{2}\left( \frac{\partial^2 E_{cell}}{\partial V^2 }\right)_{V=V_0}$. Taylor expansion is justified by the fact that our simulations and experiments use particles significantly smaller than the cell (at least 10-20 times in volume).
Finally, in our model ligand-receptor binding is not described explicitly at the molecular scale, but is accounted for by a membrane-particle contact potential $V_{LR}(\mathbf{r})$, where $V_{LR}(\mathbf{r})=-V_{LR}^{0}$ if a membrane patch is within a distance $R_0$ of the particle and zero if further away. Specifically, the associated energy is given by
\begin{equation}
E_{\text{LR}}=\int_{m} d^{2}\mathbf{r}\text{ }V_{\text{LR}}(\mathbf{r}) \label{eq:free_energy_LR}\text{,}
\end{equation}
where the integral is performed over the cell-membrane area. Effectively, $V_{LR}^{0}$ is given by the product of the individual ligand-receptor binding energy and the density of ligand-receptor bonds, divided by the density of vertices on the model membrane (see below). The width of the square potential $R_0$ is chosen to be very small compared to the other length-scales of the model $R_0<0.1R$ and does not influence the results.
\newline

\noindent {\bf Finite-element approximation.} 
Simulations of phagocytic engulfment were implemented by discretizing the cell and particle surfaces using the Surface Evolver software \cite{brakke}. This software is designed to perform energy minimization on flexible surfaces, and is freely available from http://www.susqu.edu/facstaff/\ b/brakke/evolver. The software includes a built-in programming language, which we used to implement a Monte Carlo algorithm (see below). The cell membrane is approximated by a finite number of vertices, used to create a triangular mesh. The software computes the local energy density at each vertex and sums up the energy contributions from all the surface elements to obtain the total free energy Eq. 1. 
\newline

\noindent {\bf Main model parameters.} 
Our model uses four tunable biophysical parameters. Unless otherwise specified, we have used the set of Standard Parameters (SP), chosen according to experimental measurements \cite{charras2005,herant2005,herant2006,zhelev1994} when possible (see Supplementary Information, section 1), but ultimately to produce realistic cup shapes (see Figure 2). This set of parameters includes: the cell membrane bending rigidity $\kappa_b$ and surface tension $\sigma$, respectively set to $1.3 \times 10^{-2}\text{ pN}\text{}\mu \text{m}$ and $6.2 \times 10^{-6}\text{ mNm}^{-1}$, {\it i.e.} slightly below the experimentally measured values since local changes in chemical composition of the cups membrane may reduce these parameters \cite{gordon1980,greenberg1999,groves2008,hallett2007,swanson2004,tardieu1992}.
The third parameter is the total binding energy density $\epsilon=58.5\text{ pN}\text{}\mu \text{m}^{-1}$. This value was estimated from measurements of the individual Fc$\gamma$R-IgG binding free energy $\Delta F_{LR} \approx 20k_{B}T$ \cite{keown1998,phillips1987,raychaudhuri1985,wallace1997}, the average density $d_{LR}=270-435\text{ }\mu$m$^{-2}$ of IgG-Fc$\gamma$R bonds \cite{gandour1983}, and the fact that in response to diffusion and trapping or signaling, receptors may cluster at the cup. Finally, the local constraint on cell-volume has been chosen $\kappa_{P}=2.56 \times 10^{-5}\text{ pN}\text{}\mu \text{m}^{-5}$ to allow 20 percent volume variation in line with observation \cite{moseley2009,tzur2009}.
\newline

\noindent {\bf Monte Carlo simulations.}
The Surface Evolver was only used to obtain a triangular mesh (vertices connected by edges) of the cell membrane, and to resample the membrane as the uptake progresses. The cell-membrane evolution was implemented using finite-temperature Monte Carlo Metropolis simulations \cite{metropolis1953,piotto2004}. Details of the simulation can be found in the Supplementary Information, section 2. Briefly, our algorithm calculates the total energy of the initial membrane configuration, then randomly selects a point to be the center of a membrane fluctuation, and a random direction and lateral extension of the fluctuation. Subsequently, the energy of the new membrane configuration is calculated, and compared to the initial energy. If the membrane fluctuation decreased the energy, the trial fluctuation is accepted and the procedure is reiterated starting from the new membrane configuration. To the contrary, if the membrane energy increased with the trial fluctuation, the latter may be rejected with some probability depending on the configuration's energy difference. In this case, a new fluctuation is attempted from the initial configuration. Between trial fluctuations, the cell-membrane vertices are examined. For the active zipper, the vertices within the closed neighborhood of the particle are immobilized for the remainder of the simulation. For the passive zipper, every membrane fluctuation may be reversed at a later time. 
\newline

\noindent {\bf Additional simulation parameters.} 
(1) The amplitude of a membrane fluctuation. This parameter is set to  $0.5R_0$ in all the simulations. (2) The mesh size, describing the maximal distance between two neighboring vertices. This parameter is set to $R_0$. (3) Minimal width of a membrane fluctuation. To ensure that a minimum number of vertices is involved in each fluctuation, this parameter is set to $4R_0$. (4) Mesh refinement range. Our simulation script automatically refines the mesh locally around a previously immobilized vertex within this range. This parameter is set to the particle radius $R$, leading to reasonably smooth cup shapes in a reasonably short calculation time (1-2 days for complete uptake using a Intel(R) Core(TM)2 Quad  CPU working at 2.50GHz and run by the RedHat EL5 linux distribution with 4GB RAM).

\subsection*{Experimental techniques}

\noindent {\bf Cells, plasmids and antibodies.} COS-7 fibroblast cells were obtained from American Type Culture Collection (ATCC) and cultured in Dulbecco's Modified Eagle's Medium (DMEM), supplemented with 10\% foetal bovine serum (FBS) and penicillin/streptomycin (Invitrogen). cDNAs encoding wild-type and Y282F/Y298F human Fc$\gamma$RIIa from pRK5-Fc$\gamma$RIIa \cite{caron98} and pRK5-Y282F/Y298F-Fc$\gamma$RIIa \cite{cougoule06} were subcloned into pEGFP-N1 (Clontech) using primers 5'-ggtccaactgcacctcggt-3' and 5'-ccccccgaattctgttattactgttgacatggtc-3'. The cytoplasmic tail truncation mutant 239-Fc$\gamma$RIIa was generated from the pRK5-Fc$\gamma$RIIa template using primers 5'-ggtccaactgcacctcggt-3' and 5'-gggggggaattctcctgcagtagatcaaggccact-3'. Rabbit anti-bovine serum albumin (BSA) serum was purchased from Sigma-Aldrich. Alexa-conjugated secondary antibodies and phalloidin were purchased from Invitrogen.
\newline

\noindent {\bf Transfection and phagocytic challenge.} COS-7 cells were transfected with GFP-tagged Fc$\gamma$RIIa constructs using an Amaxa Nucleofector and Nucleofector cell line kit R following the manufacturer instructions. For phagocytosis assays, transfected cells were seeded onto glass coverslips in 24-well plates at a density of 15,000 cells/coverslip and incubated at 37$\,^{\circ}$C for 72 h. 1 hour before commencement of phagocytosis assays, cells were incubated for 1 hour at 37$\,^{\circ}$C with serum-free DMEM plus 10 mM Hepes (Invitrogen). $1.5\mu$m- and $3\mu$m-radius latex-polystyrene particles (Sigma-Aldrich) were opsonized by first incubating overnight at 4$\,^{\circ}$C with 3\% BSA fraction V in PBS (Sigma-Aldrich) followed by incubation with 1:100 dilution of rabbit anti-BSA in PBS for 1 hour at room temperature. Particles were re-suspended in ice-cold serum-free DMEM plus 10 mM Hepes at a concentration of 1.5x10$^6$ particles/ml and 500 $\mu$l added to each coverslip.  Plates were incubated on ice for 10 minutes to allow binding of particles. Medium was then replaced with pre-warmed serum-free DMEM plus 10 mM Hepes and plates were incubated at 37$\,^{\circ}$C, then processed for scoring or microscopy as described below.  Experiments carried out on cells treated with cytochalasin D were conducted as above with the addition of a 20 min pre-incubation step with 0.2 $\mu$M cytochalasin D in serum-free DMEM plus 10 mM Hepes at 37$\,^{\circ}$C immediately before incubation of cells with opsonized particles. This concentration of cytochalasin D was included in all further incubation steps.
\newline

\noindent {\bf Imaging of phagocytic cups.} Plates were placed on ice after a 20 minute incubation at 37$\,^{\circ}$C and medium replaced with a 1:500 dilution of anti-rabbit Alexa 488 in 3\% BSA/PBS at 4$\,^{\circ}$C for 5 min. Cells were fixed after incubation at 37$\,^{\circ}$C for the appropriate amount of time with ice-cold 4\% paraformaldehyde/PBS, permeabilised and labelled with goat anti-rabbit Alexa 633, and phalloidin Alexa 555 for visualizing F-actin at room temperature for 30 min. Z-series image stacks were acquired on a Zeiss LSM-510 confocal microscope using a step size of 0.4 $\mu$m.
\newline

\noindent {\bf Image analysis of phagocytic cups.} 
Two fluorescence channels (IgG, Fc$\gamma$R-GFP) were acquired and analyzed using MATLAB (MathWorks). During the acquisition process, the Fc$\gamma$R-GFP fluorescence intensity was set to zero at any pixel where the IgG intensity is null. The fluorescence intensity distribution of IgG was used to determine the coordinates of the centers of particles with their corresponding radii, using an automated search based on the Hough transform \cite{ballard1981}, available online at www.mathworks.com/matlabcentral/fileexchange/4985 within each 2D image. The percentage of engulfed particle surface area was calculated by comparing the local Fc$\gamma$R-GFP (cell membrane) and IgG (particle surface) intensity distributions within a sphere $S_0$ of radius $4R/3$ whose center coincides with the particle center. 
Two methods were used to quantify the variability of the cups. The method used in the main text cuts the three-dimensional image (and hence the circular projections of particles within each imaging plane) in twenty-four angular segments, and finds the highest plane in which Fc$\gamma$R-GFP fluorescence intensity is detected in the immediate neighborhood of the particle for each segment (see Figure 4). This analysis produces a distribution of membrane height {\it versus} angular segment index, for which we compute the average and the mean-square deviation. The average height reached by cell membrane is roughly proportional to the surface engulfed (see Supplementary Fig. S10), and the mean-square deviation divided by the square root of the average height quantifies the variability of the phagocytic cup, excluding a trivial size dependence on the variability. The less accurate alternative method determines the angular distribution of Fc$\gamma$R-GFP fluorescence intensity within the particle's equator plane in the particle's immediate neighborhood, keeping only particles whose uptake level comprises between $30$ and $70\%$ (roughly half-engulfed particles). Specifically, this region is cut into twenty-four identical angular segments, for which the total Fc$\gamma$R fluorescence intensity is calculated. Then the average intensity per segment is calculated, as well as the standard deviation. The higher the standard deviation, the more variable the cup.

\section*{Accession Numbers}

The Fc$\gamma$RIIa receptor is referenced in protein database Genbank (http://www.ncbi.nlm.nih.gov/Genbank) under the accession number CAA01563. Signaling-dead mutant receptor Y282F/Y298F-Fc$\gamma$RIIa is obtained by replacing tyrosines (Y) with phenylalanines (F) at positions 282 and 298.
    
\section*{Authors contributions}
\noindent
ST participated to the conception of the model, and designed the simulation and image analysis algorithms. He carried out the simulations, image analysis, and statistical analysis of data as well as drafted the manuscript. AD and GT carried out the fluorescence imaging experiments and participated in writing the manuscript. RGE participated to the conception of the model and the interpretation of both theoretical and experimental results, and contributed to the writing of the manuscript. All authors read and approved the final manuscript.

\section*{Acknowledgements}
This paper is dedicated to the memory of Emmanuelle Caron, who tragically passed away in 2009. We acknowledge Micah Dembo and G\"unther Gerisch for helpful discussions, and Ken Brakke for help with the Surface Evolver software. We thank Vania Braga, Tony Magee and Brian Robertson for careful reading of the manuscript, and Suhail Islam for computational support. All authors would like to acknowledge funding from the Center for Integrative Systems Biology at Imperial College (CISBIC). RGE was additionally supported by the Biotechnology and Biological Sciences Research Council grant BB/G000131/1.

\end{document}


{\linespread{1.0}\small
\begin{widetext}
\begin{center}
{\Large \bf The zipper mechanism in phagocytosis: energetic requirements and variability in phagocytic cup shape \ \\ \ \\ Supplementary information}
\ \\
\ \\
\ \\
Sylvain Tollis$^{1, 2,}$, Anna E Dart$^{2,3}$, George Tzircotis$^{2, 3}$, and Robert G Endres$^{1, 2, *}$\\
\ \\
$^{1}$Division of Molecular Biosciences, Imperial College London \\
$^2$Centre for Integrated Systems Biology at Imperial College \\
$^{3}$Division of Cell and Molecular Biology, Imperial College London, \\South Kensington campus, SW7 2AZ, London, United Kingdom\\
\ \\
$^*$ To whom correspondence should be addressed. E-mail: r.endres@imperial.ac.uk
\end{center}
\ \\
\ \\
\ \\
\tableofcontents 
\setcounter{tocdepth}{4}
\ \\
\end{widetext}
\ \\
\ \\
\newpage
\ \\
\newpage
\section*{Additional discussion on parameter values and simulations}
\subsection*{1 - Choosing model parameters}
According to previous experimental studies of phagocytes, a cell's bulk bending rigidity $\kappa_b$ may vary in a range from $0.03-2\text{ }\text{pN}\text{}\mu \text{m}$ depending on the type of cell and measurement method used (see \cite{zhelev1994} and references therein). Similarly, the bulk surface tension $\sigma$ may range from $2-4 \times 10^{-2}\text{ }\text{mNm}^{-1}$ in resting neutrophils but can vary in other cell types \cite{herant2005,herant2006,zhelev1994}. However, during phagocytosis of large particles with radius $5\mu$m by aspirated spherical neutrophils, an abrupt increase in surface tension was observed (to $0.8\text{ }\text{mNm}^{-1}$ \cite{herant2006}) during the uptake. Note also that, $\kappa_{b}$ and $\sigma$ may vary on the scale of the cell \cite{charras2005,huang2005,roduit2008}. Changes in local phospholipid and protein composition of the cups membrane may influence $\kappa_{b}$ \cite{swanson2004}. Furthermore, during engulfment membrane delivery from internal buffers \cite{gordon1980,greenberg1999,groves2008,tardieu1992} and unfolding of surface membrane folds \cite{hallett2007} may strongly reduce $\sigma$ at the cup. Therefore in our simulations we use values of $\kappa_{b}$ and $\sigma$ slightly lower than previously cited values, {\it i.e.} $\kappa_{b}=1.3 \times 10^{-2}\text{ pN}\text{}\mu \text{m}$ and $\sigma=6.2 \times 10^{-6}\text{ mNm}^{-1}$ in order to produce realistic cup shapes. 
\newline
\indent
The Fc$\gamma$R-IgG binding free energy $\Delta F_{LR}$ was measured to $16k_{B}T$ \cite{phillips1987}, $18-20k_{B}T$ \cite{raychaudhuri1985,wallace1997}, and $23k_{B}T$ for IgE to Fc$\epsilon$ RI binding \cite{keown1998} at physiological temperature. The density $d_{LR}$ of IgG-Fc$\gamma$R bonds was estimated to range from $270-435\text{ }\mu$m$^{-2}$ for macrophages \cite{gandour1983}. The total binding energy density $\epsilon$ is consequently of the order of $\Delta F_{LR}d_{LR}=30\text{ pN}\text{}\mu \text{m}^{-1}$. Note that due to simple diffusion or in response to signaling, additional receptors may be recruited to the phagocytic cup, which may lead to a higher binding energy density. Additionally, receptors may cluster, leading to a non-uniform ligand-receptor binding-energy density in cells. Taken together, we chose a ligand-receptor binding energy density $\epsilon=58.5\text{ pN}\text{}\mu \text{m}^{-1}$ slightly higher than measured in resting cells.
\newline
\indent
\begin{figure}
  \centering
    \includegraphics[]{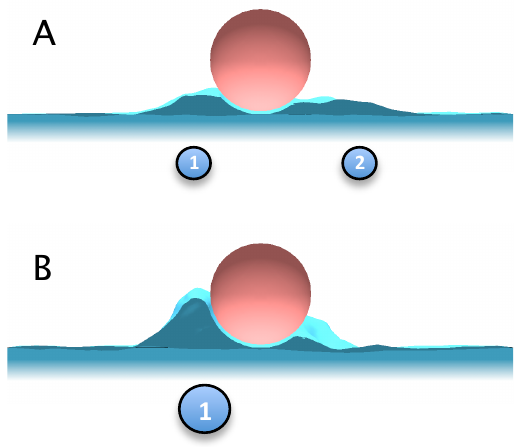}
\caption{\label{fig01-SI} Ratchet model in action. Panels (A) and (B) show phagocytic uptake at two different times with $t_{A}<t_{B}$ for the same simulation. (A) The fluctuation (1) occurs close to the particle, whereas the fluctuation (2) does not bring the cell membrane in contact with the particle. Panel (B) shows that the membrane at position (1) has progressed around the particle, whereas the membrane at position (2) has almost completely retracted to its initial flat state, demonstrating the importance of membrane immobilization by effective actin polymerization during uptake.}
\end{figure}
\begin{figure*}
  \centering
    \includegraphics[]{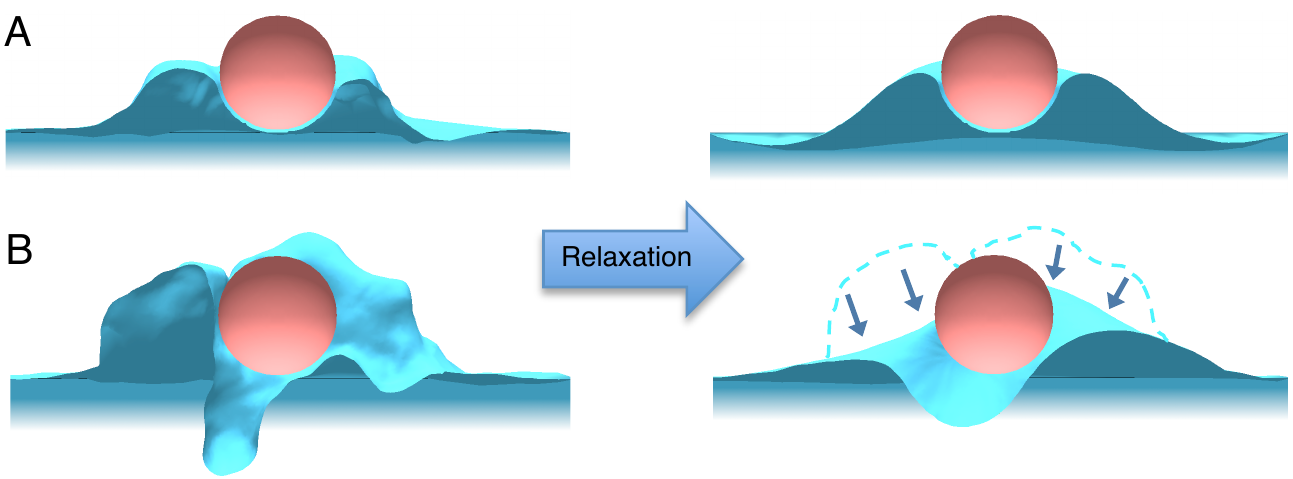}
\caption{\label{fig07-SI} Relaxation of phagocytic cup to the {\it ground state}. (A) Phagocytic cup shape at half-engulfment obtained for the active zipper with the set of Standard Parameters ({\it left}); cup relaxed at zero absolute temperature ({\it right}). (B) Phagocytic cup shape at half-engulfment obtained for the passive zipper with standard set of parameters ({\it left}); cup relaxed at zero absolute temperature ({\it right}), leading to unzipping.}
\end{figure*} 
\begin{figure}
  \centering
    \includegraphics[]{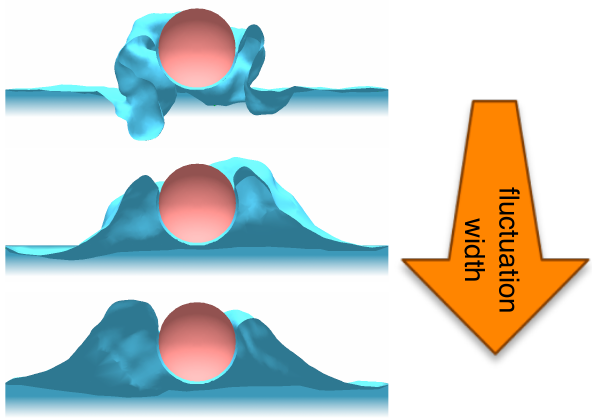}
\caption{\label{fig08-SI} Influence of the maximal width of membrane fluctuations on cup shape. Shown are ({\it top}) narrow fluctuations with maximal width $w_{\text{max}}=1.5R$, ({\it middle}) medium wide fluctuations with $w_{\text{max}}=3R$, and ({\it bottom}) wide fluctuations with $w_{\text{max}}=4R$. Parameter $R$ is the particle radius.} 
\end{figure}
The cell-volume constraint was chosen $\kappa_{P}=2.56 \times 10^{-5}\text{ pN}\text{}\mu \text{m}^{-5}$. This reflects strong regulation of cell volume on the one hand \cite{morris2001,wehner2003}, but, on the other hand, allows volume fluctuations of about 20\%, in line with observed distributions of cell volumes \cite{moseley2009,tzur2009}.
\newline
\indent
\begin{figure}
  \centering
    \includegraphics[]{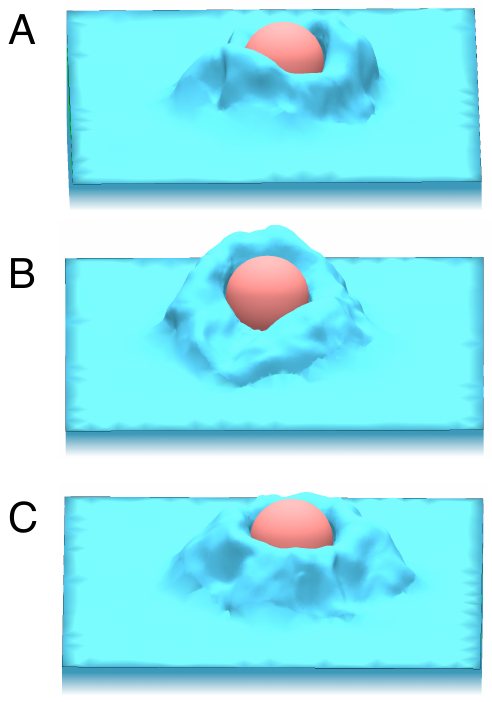}
\caption{\label{fig03-SI} Reproductibility of cup shapes for simulations under the same conditions. Side views of three simulations with the set of Standard Parameters (for approximately 50\% engulfment).}
\end{figure} 
Finally, the algorithm convergence is maximized if approximately half of the trial moves are accepted. To satisfy this requirement without constraining the biophysical parameters, we have modulated the trial moves acceptance rate by choosing an intermediate cutoff $w_{\text{max}}$ for the width of the membrane fluctuations (see below). The above listed parameter values for $\kappa_{b}$, $\sigma$, $\epsilon$ and $\kappa_{P}$ constitute the standard set of parameters values.

\subsection*{2 - Details of Monte Carlo algorithm}
\begin{figure*}
  \centering
    \includegraphics[]{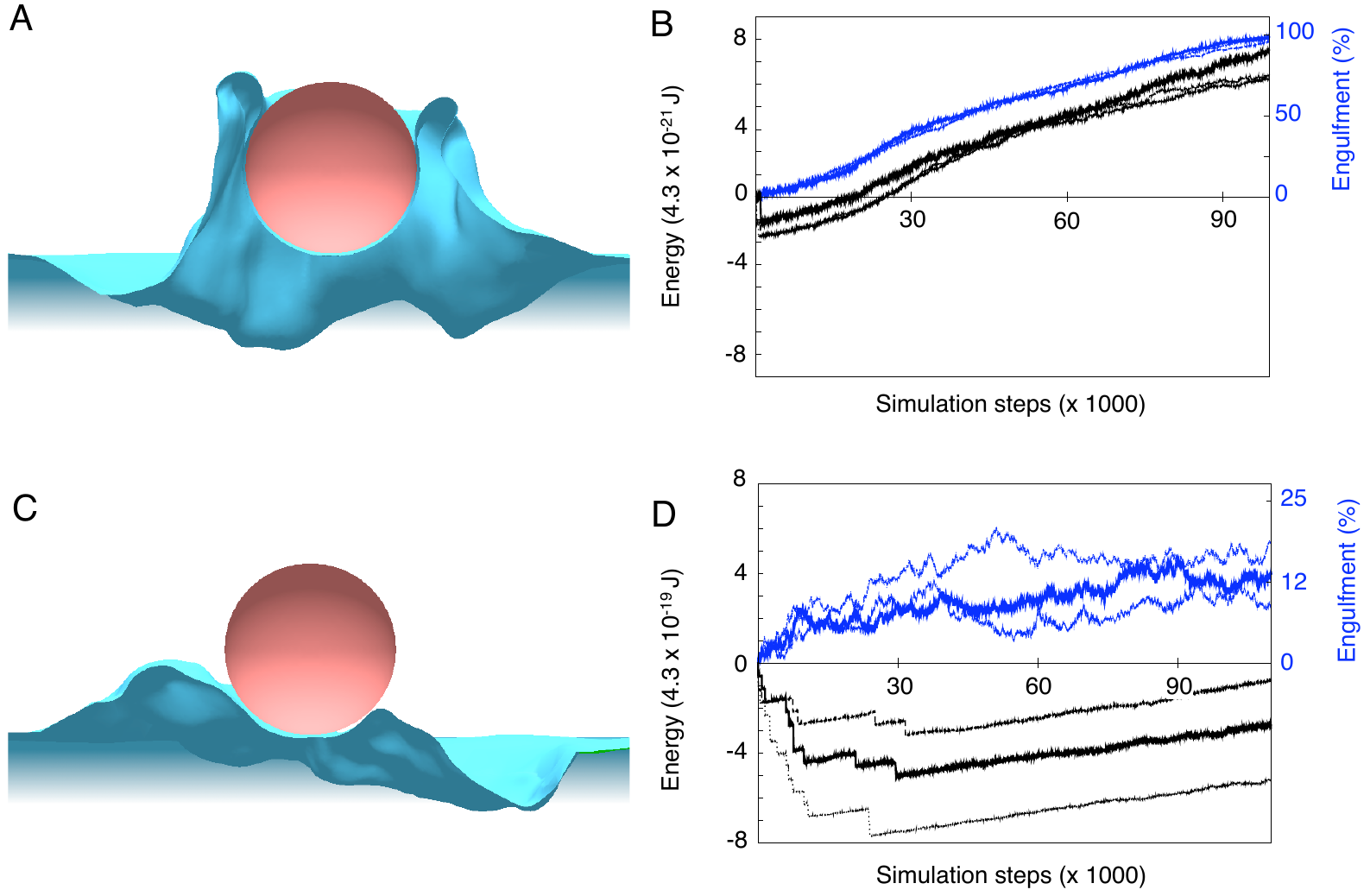}
\caption{\label{fig11-SI} Active versus passive engulfment of a large particle. (A) Cross section of phagocytic cup obtained for the active zipper. (B) Time course of total energy (black) and percentage of engulfment (blue). (C) Cross section of phagocytic cup obtained for the passive zipper (same overall simulation time). (D) Time course of total energy (black) and engulfment (blue) for the passive zipper. The particle radius is $R=3\mu$m. Three simulations were carried out in (B,D).}
\end{figure*}
Within the Surface Evolver, we implemented finite-temperature simulations using a Monte Carlo Metropolis algorithm \cite{metropolis1953,piotto2004} to describe the membrane dynamics. Our Monte Carlo algorithm includes four steps: 
\newline
(1) Calculation of the total energy of the initial membrane configuration $E_1$. (2) Random selection of a vertex $i_0$ for a membrane fluctuation (trial move) and a vector $\mathbf{u}$ for the direction of the trial move. The lateral width $w$ of the trial move is randomly chosen between $0$ and an upper limit $w_{\text{max}}$, possibly determined by the spacing between actin cortex-membrane linkage proteins \cite{charras2006}. The trial move consists in moving the vertex $i_0$ and all the vertices within a distance $d<w$ from $i_0$ by the vector $0.5\left(1+\cos(d/w)\right) \mathbf{u}$. This particular function was chosen because of its regular properties at $d=w$ (vanishing of the function and its first derivative). (3) Calculation of the total energy $E_2$ after the trial move. (4) Decision to whether accept or reject the trial move is based on the Metropolis criterion. If $E_2 \leq E_1$, the move is accepted because the new membrane configuration has a lower energy than the initial one. If $E_2 > E_1$, the move is accepted with a probability $\exp \left[-(E_2-E_1)/(k_{B}T)\right]$, where $T$ is the simulation temperature in Kelvin. If the trial move is rejected, a new trial move is selected from the same initial configuration; if it is accepted a new trial move is selected from the new membrane configuration. Between each trial move, the cell membrane vertices are examined. For the active zipper, the vertices within a cutoff distance from the particle (chosen $0.5R_0$) are immobilized, {\it i.e.} these vertices do not move anymore. As a consequence, the corresponding patch of membrane is stabilized, and the ligand-receptor bonds are irreversible. 
\newline
\indent
\begin{figure*}
  \centering
    \includegraphics[]{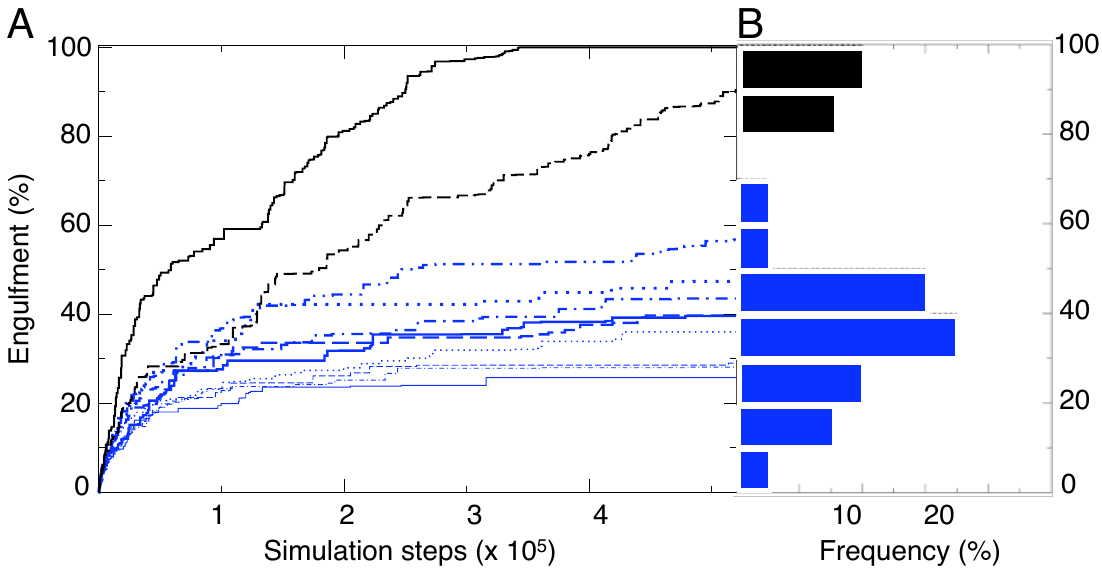}
\caption{\label{fig12-SI} Stalling at partial engulfment. (A) Time course of the percentage of engulfment as a function of increasing surface tension in simulations. Shown are completed (or nearly completed) uptake (black lines) and partial (stalled or very slowly progressing) uptake (blue lines). Surface tension values ($\sigma \times 10^{-4}\text{ mNm}^{-1}$) used are $0.8$ (black solid), $1.1$ (black dashed), $1.4$ (thick blue dash-double dotted), $1.6$ (thick blue dotted), $1.9$ (thick blue dashed-dotted), $2.1$ (thick blue dashed), $2.3$ (thick blue solid), $2.8$ (thin blue dotted), $3.1$ (thin blue dash-dotted), $3.4$ (thin blue dashed), and $3.7$ (thin blue solid). (B) Distribution of engulfed particle surface area measured in our experiments for COS-7 cells transfected with wild-type receptors (WT-Fc$\gamma$R) after 10min. Black and blue bars correspond to the two peaks of the bimodal distribution.}
\end{figure*}
For the passive zipper, none of the vertices are immobilized and all the trial moves may be reversed at a later time. Note that our choices of {\it simulation} parameters imply that the number of accepted and rejected fluctuations are of the same order of magnitude, which improves the convergence of the Metropolis algorithm. 
\newline
\indent
\begin{figure}
  \centering
    \includegraphics[]{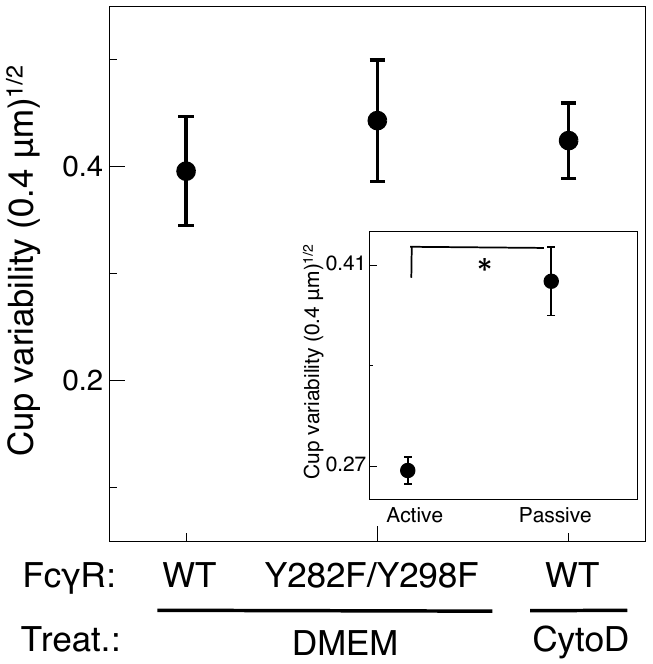}
\caption{\label{fig04-SI} Statistical analysis of phagocytic cup shapes. Cup variability (for 60 to 100\% engulfed particles) was determined for COS-7 cells transfected with wild-type Fc$\gamma$R (WT-Fc$\gamma$R) or signaling-dead mutant (Y282F/Y298F-Fc$\gamma$R), as well as WT-Fc$\gamma$R cells treated with $0.2\mu$M cytochalasin D (WT-Fc$\gamma$R$+$CytoD). The data points represent the variability averaged over all the phagocytic cups measured for given condition and range of uptake. Error bars represent the statistical standard error of the cup variability. ({\it Insets}) Theoretical cup variability calculated from corresponding model simulations. Student's t-test (*): p-value is less than 0.0001.}
\end{figure}
Our equilibrium Monte Carlo approach is justified {\it a posteriori} to study the dynamics of uptake, since only a very small proportion (less than 1\%) of the accepted trial moves is irreversibly immobilized. Hence, the algorithm samples membrane configurations extensively in line with equilibrium thermodynamics.

\section*{Additional simulations}
\subsection*{3 - Ratchet model in action}
Our ratchet-like biophysical model is based on effective actin polymerization to render ligand-receptor bonds irreversible. As ligand-receptor binding and hence signaling can only occur in the immediate neighborhood of the particle, random membrane fluctuations far from the particle may retract with increasing simulation time. {\bf Figure S1} gives an example of a reinforced membrane fluctuation near the particle and a vanishing membrane fluctuation further away from the particle, illustrating how this concept is implemented in the simulations.

\subsection*{4 - Role of engulfment kinetics in cup shape}
\begin{figure}
  \centering
    \includegraphics[]{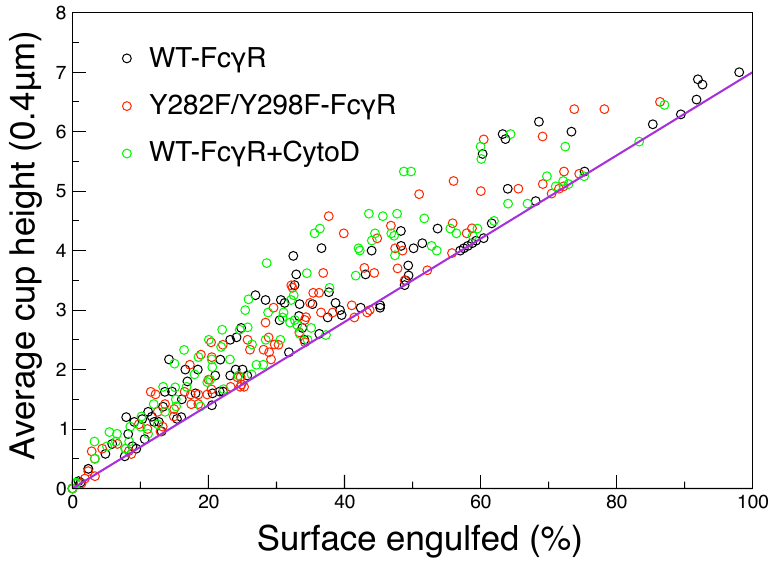}
\caption{\label{fig05-SI} Average cup height $<\!h\!>$ versus percentage of engulfed particle surface area $S$. Each circle corresponds to an imaged cup (black for WT-Fc$\gamma$R, green for Y282F/Y298F-Fc$\gamma$R, and red for WT-Fc$\gamma$R$+$CytoD). For a perfectly regular cup, both quantities are related by $S(\%)=100<\!h\!>(\mu \text{m})/(7 \times 0.4)$. Due to shape variability and finite number of angular segments used, values obtained lie slightly above the diagonal straight line.} 
\end{figure}
Our simulations show that cup shape depends on the kinetics of phagocytic uptake. To demonstrate this we relaxed a phagocytic cup obtained with a finite-temperature Monte Carlo simulation towards the {\it ground state}, {\it i.e.} the shape at zero absolute temperature in the absence of thermal fluctuations (see {\bf Figure S2}). During relaxation, the membrane evolves towards the minimal energy configuration ({\it right}) starting from the initial configuration ({\it left}). In the case of the active zipper (panel A), the resulting cup not only smoothes, but also changes shape (here it becomes broader). For other parameter choices, simulated cups are either broader or thinner than predicted by their {\it ground state} due to finite temperature stochastic engulfment, emphasizing the importance of the randomness of membrane fluctuations in shaping the cup.
\newline
\indent
\begin{figure*}
  \centering
    \includegraphics[]{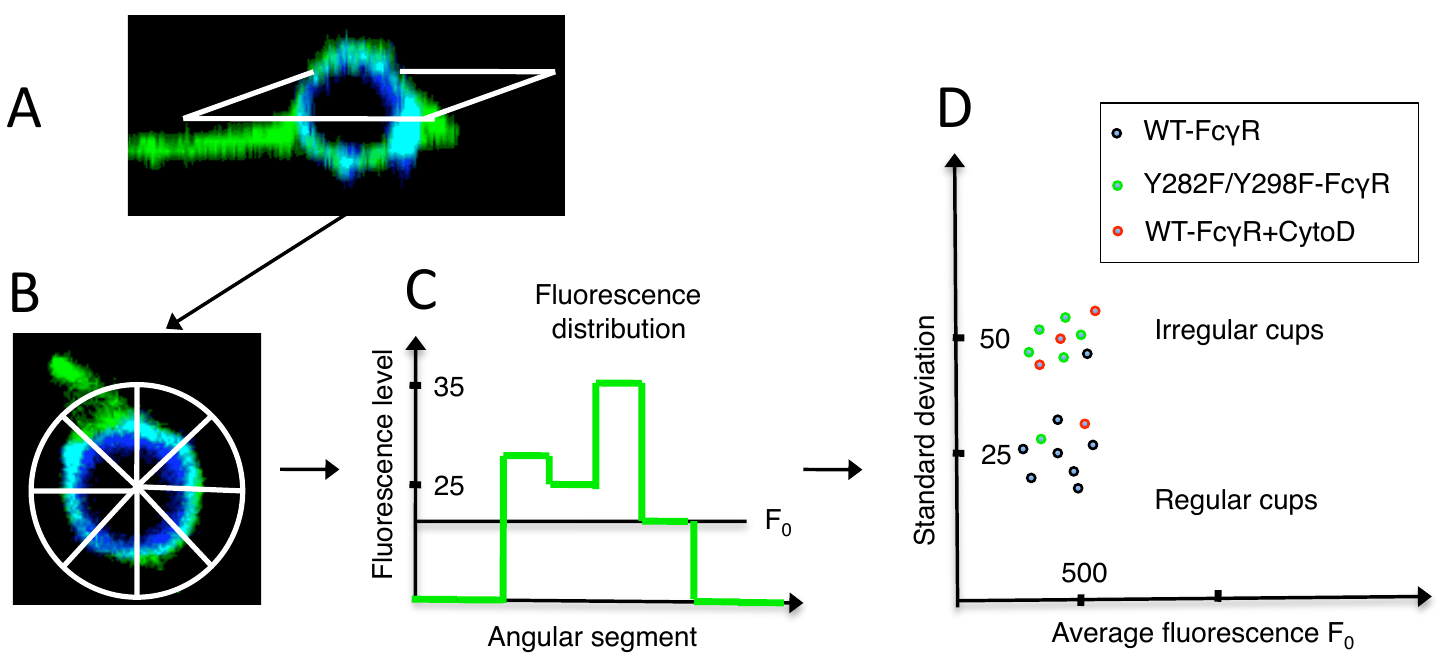}
\caption{\label{fig06-SI} Alternative image analysis of phagocytic cups. (A) Side view of a typical 3-dimensional fluorescence image, reconstructed from confocal microscopy data. Shown are the Fc$\gamma$-GFP receptors (green) and IgG ligand coating the particle (blue). The analysis plane (white square) intersects the particle at the equator. (B) Top view of the analysis plane. The green light distribution is determined using the total green light intensity in the different angular segments. (C) An example of such a distribution for one particular particle, used for extracting the average intensity $I_0$ and the mean-square deviation $\delta I$. (D) Illustrative scatter plot and classification of particles with respect to their engulfment statistics $I_0$ (horizontal axis) and normalized root mean-square deviation $\delta I_\text{N}=\delta I/\sqrt{I_0}$ (vertical axis). Regular cups correspond to a smaller root mean-square deviation of the fluorescence distribution than variable cups.}
\end{figure*}
In the case of the passive zipper (panel B), finite temperature stochastic processes are even more crucial for engulfment. Indeed, {\bf Figure S2B} ({\it left}) is obtained with a simulation of the passive zipper at finite temperature. After some time, the simulation is stopped and restarted at zero absolute temperature for relaxation of the partially engulfed state towards the {\it ground state}. After partial relaxation, the cup presented in {\bf Figure S2B} ({\it right}) is obtained. The membrane smoothes similar to the active zipper, but also retracts, resulting in unzipping. This indicates that the release of ligand-receptor binding energy is not sufficient for uptake by the passive zipper. Additional thermal membrane fluctuations are also required. Further relaxation eventually leads to a total retraction of the cell membrane for the passive zipper.

\subsection*{5 - Influence of membrane fluctuation width on cup shape}
The maximal width of the fluctuations $w_{\text{max}}$ is chosen to optimize the convergence of the Monte-Carlo Metropolis algorithm (see {\it Methods} in the main text). This algorithm accurately samples the probability distribution of the membrane configuration, and consequently of cup shapes at different stages of uptake. The convergence of the algorithm is optimal if about half of the randomly selected membrane fluctuations (trial moves) are accepted. This condition is fulfilled for intermediate values of $w_{\text{max}}$. {\bf Figure S3} shows how the choice of $w_{\text{max}}$ influences the cup shape. If $w_{\text{max}}$ is small, only laterally small membrane fluctuations are generated by the algorithm. These fluctuations do not change the surface area and cell volume significantly, and consequently are frequently accepted by the algorithm. Hence, the cup grows fast and thin ({\it top} panel). In contrast, if $w_{\text{max}}$ is large, laterally wide fluctuations may be selected, and are less frequently accepted since they can lead to drastic changes is cell volume and/or area. The fluctuation-acceptance rate is decreased, the cup grows more slowly and becomes broad ({\it bottom} panel).

\subsection*{6 - Reproductibility of cup shapes}
Phagocytic cup shape is variable, in particular without the support of the actin cytoskeleton for the passive zipper. However, even for the active zipper, engulfment is inherently stochastic due to random membrane fluctuations. As depicted in {\bf Figure S4}, simulations with the same set of parameters lead to slightly different cups after the same simulation time. This illustrates particle-to-particle variation.
\begin{figure}[!h]
  \centering
    \includegraphics[]{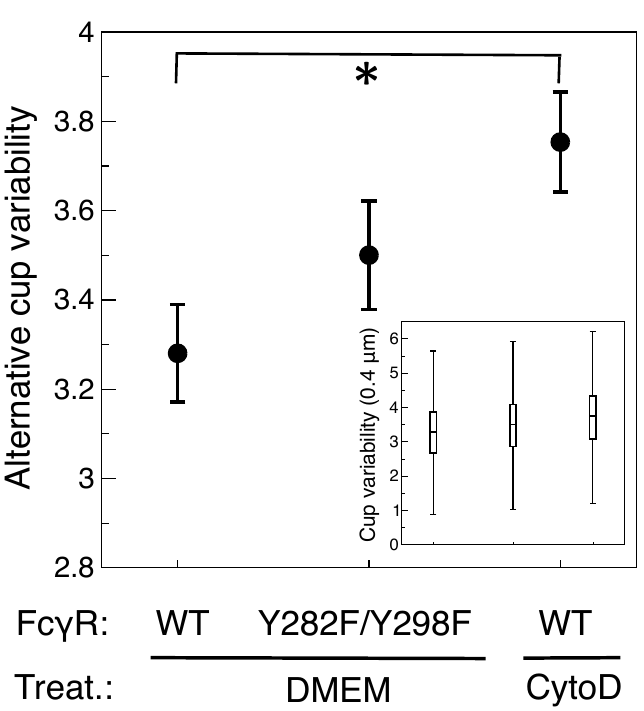}
\caption{\label{fig10-SI} Alternative statistical analysis of phagocytic cup shapes. Cup alternative variability (arbitrary units) for 30 to 70\% engulfed particles, for conditions WT-Fc$\gamma$R, Y282F/Y298F-Fc$\gamma$R, and WT-Fc$\gamma$R$+$CytoD. The data points represent the variability averaged over all the phagocytic cups measured for given condition and range of uptake. Error bars represent the statistical standard error of the cup variability. Student's t-test: p-values is 0.003 (*). ({\it Insets}) Whisker plot representation of the cup-shape alternative variability.}
\end{figure}

\subsection*{7 - Active versus passive engulfment of large particles}

In the main text we demonstrate that the passive zipper is able to take up small particles of radius $1.5\mu$m, although slower and in a more variable fashion than the active zipper. Here, we compare the ability of the two zippers to take up a larger particle of radius $3\mu$m. {\bf Figure S5} shows that the passive zipper only engulfs a few percent of the particle surface area (panels C and D), whereas the active zipper completes uptake (panels A and B). This demonstrates that, for the uptake of large particles, ligand-receptor binding in conjunction with thermal fluctuations is not sufficient, and that the functionality of the actin network is fully required.

\subsection*{8 - Stalling at partial engulfment}

A model previously published by one of the authors uses the force generated by actin on the cell membrane as a parameter, and predicts that slightly different force strengths lead to very different behaviors of the phagocytic cup \cite{tzircotis2009}. Specifically, a bimodal distribution of phagocytic cups was found, {\it i.e.} cups either fully take up the particle or stall before 50\% engulfment. This was interpreted as a mechanical bottleneck at half-engulfment (see main text). Our simulations also exhibit this effect. {\bf Figure S6A} shows how slight (much less than an order of magnitude) variations in biophysical parameters (here surface tension) may have drastic effects on the completion of uptake. This high sensitivity to small parameter changes, for specific parameter ranges, explains the particle-to-particle variation observed in our experiments ({\bf Figure S6B}).

\section*{Additional analysis of fluorescence images}
\subsection*{9 - Statistics of phagocytic cup shapes}
In Figure 5 of the main text, we demonstrate that the cup variability in experiments is increased when signaling to the actin cytoskeleton is interrupted or when cells are treated with cytochalasin D. More precisely, this result holds at the beginning of uptake between 20 and 40\% of engulfed particle area. Between 40 and 60\% of the uptake, there is no significant difference between the three experimental conditions. For completeness, we report here the cup variability between 60 and 100\% of uptake.  {\bf Figure S7} shows that almost completed cups are more regular (the variability is reduced by a factor of 2 compared to the range 40 to 60\%), and we observe no statistically significant difference between the three different conditions. Note that the variability obtained in simulations (see {\bf Figure S7} {\it inset}) is also reduced by a factor of 2, and that the relative difference in variability between the active and passive zipper is also reduced, as expected since fully completed cups are necessarily regular.

\subsection*{10 - Average cup height versus percentage of engulfed particle surface area}

Our image analysis method makes use of two different measures of uptake. The height  $h$ reached by the cell membrane (measured by confocal slice index), and the percentage of particle surface $S$ covered by cell membrane. The height is used to calculate the cup variability, and the surface engulfed is used to classify the cups (20 to 40\%, 40 to 60\%, and 60 to 100\%). {\bf Figure S8} shows that these two quantities are indeed highly correlated. For a perfectly regular cup, the height $<\!h\!>$, averaged over angular segments, is given by $<\!h\!>=7S/100 \times 0.4\mu$m, because each particle is imaged in $7$ confocal planes. As shown, this relation does not exactly hold for variable cups, and the average height is slightly higher than predicted by the engulfed surface area. This minor deviation is due to the finite number of angular segments used. 

\subsection*{11 - Alternative image analysis of phagocytic cup shapes}

To rule out any bias in the analysis, we additionally use an alternative method to characterize cup variability. The method is based on the observation that a perfectly regular cup (at half-engulfment) reaches the equator plane all around the particle circumference, corresponding to an engulfed surface area $S=50\%$, and an average height $<\!h\!>=1.5\mu$m. For such a cup the distribution of Fc$\gamma$R-GFP fluorescence signal (indicating the presence of the cell membrane) should be uniform around the projection of the particle onto the equator plane (circle of radius $R$). In contrast, a variable half-engulfed cup has an irregular distribution of fluorescence intensity as the particle is only partially covered by cell membrane in its equator plane. To exploit this observation, we calculate the distribution of Fc$\gamma$R-GFP fluorescence along the particle's perimeter in angular segments. The standard deviation of this distribution quantifies the alternative measure of the cup variability, illustrated in {\bf Figure S9}.
\newline
\indent
{\bf Figure S10} shows the result from the statistical analysis of phagocytic cup shapes as obtained with the alternative method. Although this analysis method is more restrictive and inaccurate, it confirms our result from the main text. Using 30 to 70\% engulfed particles we obtain that cells transfected with wild-type Fc$\gamma$R produce significantly less variable cups than cytochalasin D treated cells (Student's t-test: p-value$=0.003$).

